\documentclass[sn-nature]{sn-jnl}
\usepackage{graphicx}
\usepackage{multirow}
\usepackage{amsmath,amssymb,amsfonts}
\usepackage{amsthm}
\usepackage{mathrsfs}
\usepackage{xcolor}
\usepackage{textcomp}
\usepackage{manyfoot}
\usepackage{booktabs}
\usepackage{algorithm}
\usepackage{algorithmicx}
\usepackage{algpseudocode}
\usepackage{listings}
\usepackage{amsmath}
\usepackage{hyperref}
\usepackage{caption}

\usepackage{todonotes}
\usepackage{tikz}
\usepackage{lipsum}
\usepackage{graphicx}
\usepackage{dcolumn}
\usepackage{bm}
\usepackage{verbatim}
\usepackage{tikz}
\usetikzlibrary{quantikz2}
\usepackage{multirow}
\usepackage{amsmath}
\usepackage{amssymb}
\newcommand{\sign}{\text{sign}}

\begin{document}
\title{Explainable quantum regression algorithm with encoded data structure}
\author*[1]{\sur{C.-C. Joseph Wang}}\email{josephwang13@gmail.com}
\affil*[1]{\orgdiv{Quantum Computational Science Group, Quantum Information Science Section}, \orgname{Oak Ridge National Laboratory}, \orgaddress{\city{Oak Ridge}, \postcode{37831}, \state{Tennessee}, \country{USA}}}
\author[2]{\sur{F. Perkkola}}
\author[2]{\sur{I. Salmenper\"{a}}}
\author[2]{\sur{A. Meijer-van de Griend}}
\author[2]{\sur{J. K. Nurminen}}
\affil[2]{\orgdiv{Department of Computer Science}, \orgname{University of Helsinki}, \state{Helsinki}, \country{Finland}}

\abstract{
Hybrid variational quantum algorithms are promising for solving practical problems in application areas, which include quantum machine learning, combinatorial optimization, and quantum chemistry simulation on noisy quantum computers. However, variational quantum algorithms (currently derived from randomized hardware-efficient ansätze or adaptive ansätze) become a black box, not trustworthy for model interpretation, and not to mention application deployment in informing critical decisions. In this paper, we construct the first interpretable quantum regression algorithm inherently, in which the quantum state exactly encodes the classical data table and the variational parameters correspond directly to the regression coefficients, which are real numbers by construction, providing a high degree of model interpretability and minimal cost to optimize due to the right expressiveness. We also exploit the encoded data structure to reduce the gate complexity of computing the regression map. To reduce circuit depth in nonlinear regression, our algorithm can be extended by directly constructing nonlinear features via classical preprocessing, such as independent encoded column vectors. By design, the model performance is determined by the cost function measurement results synchronous to the mean squared errors (MSE) for the regression models. We derive the read-out errors induced by one-hot encoding and compact encoding; the required physical qubit resources are exponentially compressed for the compact encoding to be favorable for noisy quantum devices. We also derive explicitly the data-aware sample complexity $\in \mathcal{O}\left(\sigma^{2} \ln (1/\alpha)/\epsilon^{2}\right)$
per classical optimization iteration under the error budget $\epsilon$  and confidence tolerance $\alpha$ for further empirical benchmarks.
}

\maketitle
\section{Introduction}\label{sec1}
The interpretability and explainability of predictive models are essential~\cite{Classical_ML} for the wider adoption of machine learning and artificial intelligence applications, especially in domains where faulty model interpretation can have serious consequences. For example, in healthcare and financial applications, strict regulations require models with clear interpretation to validate model predictions for the model to be approved/trusted. Model interpretability is an equally valid criterion for quantum machine learning, but so far, it has received little or no attention. Although a quantum regression algorithm was proposed a decade ago~\cite{quantum-algorithm-for-data-fitting, prediction-by-linear-regression-on-a-quantum-computer,fast-quantum-algorithms-for-least-squares-regression} for its fundamental importance, and other approaches based on matrix inversion and quantum kernel methods have been proposed recently~\cite{Somma, Paine}, these works assumed fault-tolerant (no noise effects) quantum hardware and did not address model interpretation issues. In recent work with a hybrid variational ansatz to mitigate hardware noise~\cite{LANL}, as we did, they did not consider interpretation values for potential quantum applications.
We approach quantum regression from a variational perspective with a known encoded data structure and develop an algorithm that provides interpretive value and prediction power as required, useful in the noisy intermediate-scale quantum (NISQ) era~\cite{NISQ}.

Regression models are predictive models that learn the map between a target continuous variable and predictors (attributes/input variables/features) in training.
The predictor variables can generally be transformed into continuous variables with the appropriate interpretation based on the transformation performed. Regression models are important machine learning models to study due to their wider adoption in industrial applications at scale, as opposed to more complex models such as neural networks, which typically focus on predicted results and less on the descriptive correlation between the prediction outcomes and the predictors. Additional features, such as the flexibility to model nonlinear dependencies based on domain expertise and the ability to perform relevant variable selection with regularization techniques, further enhance the utility of regression modeling in statistical machine learning.

Variational quantum algorithms are undeniably the most feasible digital quantum algorithms to date~\cite{Variational quantum algorithms, A quantum approximate optimization algorithm, An adaptive variational algorithm for exact molecular simulations on a quantum computer, Hybrid Quantum-Classical Algorithms and Quantum Error Mitigation, Variational Circuit Compiler for Quantum Error Correction}. They offer a practical solution to bypass quantum hardware noise and intricate controls while maintaining their universality for quantum computation~\cite{Universal variational quantum computation}.
However, current hybrid variational algorithms (based on quantum alternating operator ansatz) are generic and approximate toward quantum applications, and therefore, the direct connection of the variational gate parameters of direct interpretable values is often lost; we embark on the quantum regression problem differently in perspective by producing the exact regression map with the optimal gate parameters directly connected
to the regression coefficients crucial for the interpretability of the regression models.

The organization of the manuscript is as follows.
In Section~2, we motivate and introduce the abstraction of the full quantum algorithm, including amplitude encoding and regression map generation with real variational parameters, which are explainable weights in classical regression problems and measurement. 
In Section~3, we present the quantum algorithm in Pauli spin language, which is pertinent for researchers working on quantum hardware. Therefore, the gate complexity for the quantum algorithm can be analyzed naturally.
In Section~4, we simulate noisy measurement data gathered by quantum hardware with bootstrap sampling in conjunction with an ensemble of regression models to find the optimal weight parameters. We also show that regularization techniques are still a valid strategy for selecting important features in the context of the variational quantum regression algorithm, as in the classical regression algorithm.
In Section~5, we conclude with our findings.

\section{The explainable quantum regression algorithm}
\subsection{Problem statement}

Regression modeling is the task of determining the relationship between a vector of independent variables (or "features") $({\bf X}_1,\ldots, {\bf X}_M) \in \mathbb{R}^{L\times M}$ and a dependent variable (or "response") ${\bf Y} \in \mathbb{R}^{L}$ from experimental data. It is one of the most common and important tasks in science, with particular prevalence in data modeling and machine learning. Usually, the relationship is assumed to be linear in ${\bf X}_m$, ${\bf Y} = \sum_{m=1}^{M} W_m {\bf X}_m$ where $(W_1,\ldots, W_m)$ are known as regression coefficients or importance weights.
However, by treating products of independent variables as additional independent variables, linear regression can also be used to model non-linear relationships. In the typical regression scenario, one has $L$ independent observations $(y_0,\ldots,y_{L-1})$ of ${\bf Y}$ and the corresponding observations $(x_{0m},\ldots,x_{(L-1)m})$ of each variable ${\bf X}_m$. The goal of (linear) regression is to determine the coefficients $(W_1,\ldots, W_m)$ that best fit the data.

We propose a new algorithm to solve the linear regression problem using variational quantum circuits, whose parameters encode the regression coefficients in a manner that allows for interpretable values. The best regression coefficients are found by classical optimization with a penalized cost function, which furthermore helps to find the subset of the most important features. A key aspect of our approach is that the structural data is encoded directly in the amplitudes of the quantum state, and the regression coefficients are encoded directly in the parameters of the quantum circuit, which leads to optimal interpretability. We note that protocols for implementing quantum amplitude encoding are still under active research ~\cite{Quantum state preparation protocol for encoding classical data into the amplitudes of a quantum information processing register's wave function, W state, TensorFlow Quantum: Impacts of Quantum State Preparation on Quantum Machine Learning Performance}. Along with our regression algorithm, we provide several state-preparation algorithms to facilitate the implementation of regression on near-term quantum computers.

\subsection{Quantum amplitude encoding}
The first step of the algorithm is to encode the observations $y_0,\ldots,y_{L-1}$, $x_{11},\ldots,x_{LM}$ in a quantum state.
For notational convenience, we define $x_{l0} \equiv y_l$ and define the matrix $\mathbf{X}$ with elements $x_{lm}$ for $l=0,1,\ldots,L-1$ and $m=0,1,\ldots,M$.
We standardize the data by shifting and rescaling the data columns so that each column $\mathbf{X}$ has a zero mean and a variance. This ensures that our algorithm is equally sensitive to all variables for the best training. The data is then globally normalized so that $\sum_{l,m} x_{lm}^2 = 1$.
Therefore, the data can, in principle, be mapped to the amplitudes of a quantum state:
\begin{equation}
|\psi_{D}\rangle = \sum_{l,m} x_{lm}|lm\rangle
\end{equation}
where $\{|lm\rangle\}$ are the computational basis states of a quantum system that has at least $L(M+1)$ orthogonal states.
For now, we do not discuss the details of possible encoding schemes or methods for preparing $|\psi_{D}\rangle$, as this would distract from the main ideas of the algorithm.
Details for physical implementation will be discussed in Section 3.

\subsection{Mapping of regression coefficients to quantum amplitudes}

Our goal is a variational circuit whose structure reflects that of the regression problem at hand and whose output is proportional to the mean-squared error (MSE) to be minimized with the optimal regression coefficients 
\begin{equation}
\hat{W}_{m\in{\{1,\cdots,~M}\}} = \mathop{\arg \min}\limits_{W_{m}\in \mathbb{R}^{M}}  \rm{\bf MSE},
\end{equation}
in which
\begin{equation}
{\rm {\bf MSE}} := \sum_{l=0}^{L-1} ( y_l - \hat{y}_l)^2
\label{Eq:cost_function}
\end{equation}
and the expected response from the model with the learned $\hat{W}_{m}$
\begin{equation}
\hat{y}_l := \sum_{m=1}^{M} x_{lm} \hat{W}_{m}.
\end{equation}

We show first how to multiply a given feature (column of $\mathbf{X}$) by a controllable coefficient.
It will be convenient for exposition to treat the row index $l$ and column index $m$ as separate quantum degrees of freedom, $|lm\rangle = |l\rangle \otimes |m\rangle \equiv |l\rangle |m\rangle$. Consider the operator,
\begin{equation}
U^{m}(\phi)
= \boldsymbol{1} \otimes e^{-i\phi |m\rangle \langle m|}
\end{equation}
which acts as an identity ($\boldsymbol{1}$) on the row (observation) register and imparts a phase to a selected element of the column (feature) register.  It maps $|l\rangle |m\rangle$ to $e^{-i\phi} |l\rangle |m\rangle$ and leaves all other basis states unchanged. Thus, when applied to $|\psi_{D}\rangle$, it maps $x_{lm} \to e^{-i\phi} x_{lm}$ to all $l$. By extension, the sequence $\prod_{m=1}^{M} U^{m}(\phi_m)$ applies a controllable phase $\phi_m$ to each column $m$ of the data. In this case, the resulting state would be
\begin{equation}
|\psi_{D}\rangle = \sum_{l, m} x_{lm}e^{-i\phi_m}|l\rangle |m\rangle.
\end{equation}
Notice that the relation between the phase $\phi_{m}$ and the coefficient of the state $|l\rangle |m\rangle$ is not exactly what we are looking for if we were to associate the phase $\phi_{m}$ with the real regression parameters. The quantum map would not be real (up to a global phase factor) and would not be linear in $\phi_{m}$ as expected for conventional linear regression.
Furthermore, the regression coefficients should range between $(-\infty, +\infty)$, while the unique range of the dependent variable $\phi_m$ is $(-\pi,\pi)$. Based on these observations, we cannot make a direct association of the phases $\phi_{m}$ with the regression weights $\hat{W}_{m}$. However, if we engineer the circuit in a target code space to yield
\begin{equation}
|\psi_{l}\rangle  \propto \sum_{m} x_{lm} ( e^{-i\phi_m}+e^{+i\phi_m}) |l\rangle |m\rangle
\propto \sum_{m}x_{lm}\cos{\phi_m} |lm \rangle.
\end{equation}
We can identify $\hat{W}_m \propto \cos{\phi_m} \in [-1.0, 1.0] $, with the proportionality chosen to bring the weights into the required range $\in (-\infty,+\infty)$.

\subsection{Quantum regression algorithm}

To engineer this mapping of phases to regression weights, we use controlled phase gates in the form of
\begin{equation}\label{eq: regression unitary}
U_{C}^{m}(\phi)
=|0\rangle \langle 0| \otimes \boldsymbol{1} \otimes e^{i\phi_m |m\rangle\langle m|} +
|1\rangle \langle 1| \otimes \boldsymbol{1} \otimes e^{-i\phi_m |m\rangle\langle m|}
\end{equation}
which act on an ancillary qubit for control, row register, and column register, respectively.
(Note that if the hardware does not natively support such a controlled gate with symmetric phases, it can be realized as an \emph{un}controlled rotation $e^{i \phi_m}$ followed by a controlled rotation $e^{-2i\phi_m}$ equivalently.)
This gate imparts the phase $ e^{i\phi_m}$ to $|0\rangle \otimes |l\rangle \otimes |m\rangle$, and leaves the states with column index $\neq m$ unchanged.
As we now show, the transformation $x_{lm} \to \cos\phi_m x_{lm}$ can be accomplished by such controlled phase gates with a suitably prepared and measured ancillary qubit. The steps of the algorithm and the corresponding evolution of the quantum state are as follows:
\begin{enumerate}
 \item Prepare the data state $|\psi_D\rangle$:
 \begin{equation}
     |\psi_{D}\rangle = \sum_{l,m} x_{lm}|lm\rangle
 \end{equation}
 (refer to physical implementation in Section 3).

 \item Prepare an ancillary qubit in the state $\ket{+} \equiv (\ket{0} + \ket{1})/{\sqrt 2}$:
 \begin{equation}
      \longrightarrow \ket{+} \otimes  |\psi_{D}\rangle.
 \end{equation}

 \item  Apply controlled phase gates $U_{C}^{m}$ for each column $m$:
 \begin{align}
     \longrightarrow & \quad \prod_{m} U_{C}^{m}(\phi_{m})
     \left( \frac{|0\rangle + |1\rangle}{\sqrt 2} \otimes  |\psi_{D}\rangle \right) \\
     & = \frac{1}{\sqrt{2}} \sum_{l,m} \left(  e^{i\phi_{m}} |0\rangle + e^{-i\phi_{m}} |1\rangle \right) \otimes x_{lm} |lm\rangle
 \end{align}

 \item Apply a Hadamard gate to the ancillary qubit:
 \begin{align}
 \longrightarrow & \quad \frac{1}{2} \sum_{l,m} \left( e^{i\phi_{m}} (|0\rangle + |1\rangle) + e^{-i\phi_{m}} (|0\rangle-|1\rangle) \right) \otimes x_{lm} |lm\rangle \\
 &=  \sum_{l,m} \left( \cos \phi_{m} |0\rangle + i \sin \phi_{m} |1\rangle \right) \otimes x_{lm} |lm\rangle
\end{align}

 \item Project the ancillary qubit onto the state $|0\rangle$:
 \begin{equation}
    \longrightarrow \quad |\Psi_{0}\rangle = \sum_{l,m} x_{lm} \cos\phi_{m} |lm\rangle
  \end{equation}

 \item Measurement by the Hermitian operator:
   \begin{equation}\label{eq: measurement operator}
      \hat{M} = \sum_{l=0}^{L-1} \sum_{m,m'=0}^{M} |lm\rangle \langle lm'|.
   \end{equation}

\end{enumerate}

As shown in Appendix A, the expectation value $\langle \hat{M} \rangle \equiv \langle \Psi_{0}|\hat{M}|\Psi_{0}\rangle$ is
\begin{align}
 \langle \hat{M} \rangle  &= \sum_{l} \left( \sum_{m} x_{lm}{\cos\phi_m} \right)^{2}  \\
   & = (\cos\phi_0)^2 \sum_{l} \left( y_l - \sum_{m=1}^{M} x_{lm} {W}_m \right)^{2}
 \label{eqs:measurement1}
\end{align}
where we identify ${W}_m = -\cos\phi_m/\cos\phi_0$ as the regression coefficient for the feature $m$ (to be optimized for model training),
and the response variable component $y_{l}$ is, by definition, the $x_{l0}$ component. With this identification, the regression model errors in the above equation can be recognized as the MSE in Eq.~\eqref{Eq:cost_function}. This result bridges the gap between our quantum regression algorithm and the conventional regression algorithm, enabling a clear interpretation of the variational parameters as discussed in our numerical studies.
In general, only the relative sign between the feature variables and the response variable matters; we can enforce the condition $\phi_{y}:= \phi_{0} \in (\frac{\pi}{2}, \frac{3}{2}\pi)$ so that $\cos \phi_0$ is always negative and nonzero so that the regression coefficient $W_m$ is well defined. 
To be explicit, we will still keep the response rotational angle $\phi_{0}$ as a variational variable in later discussions.

\subsection{Model training and regularization}
Since the goal is to minimize the model error, the simplest approach is to take the cost function $\mathcal{C}(\mathbf{W})$ to be the mean-squared error $\rm {\bf MSE}$,
\begin{equation}
  \mathcal{C}({\bf W}) = \cos^{2}{\phi_{0}}\sum_{l = 0}^{L-1}(y_l - \sum_{m = 1}^{M} x_{lm}{W}_{m})^{2} = \langle \hat{M} \rangle
\end{equation}
where the regression weight vector ${\bf W} = (W_1,\ldots,W_m)$ is an implicit function of the circuit parameters $\boldsymbol{\phi} = (\phi_0,\ldots,\phi_m)$.
The parameter vector $\boldsymbol{\hat{\phi}}$ that minimizes the cost function yields the optimal linear regression coefficients $\hat{W}_{m} = -\cos\hat{\phi}_m / \cos\hat{\phi}_0$.

A fundamental question to ask is how sensitive the cost function is to a well-trained model versus a poorly trained model
as the null (reference) model, in which the estimated response variable is the mean of the variable.
When the data used to train the model is noise-free and perfectly correlated with the response variable, we expect the cost function of a well-trained model to be zero. The minimal number of shots
with measurement error $\delta \epsilon$ depends on the value of the cost function (see Appendix~B).

In practice, the cost function in regression models typically includes regularization terms to bias toward models that fit the data well with fewer features, which helps avoid overfitting~\cite{statistical machine learning} and mitigate the barren plateau from the important feature truncation from the regularization. The cost function is then modified to the following:
\begin{equation}
\mathcal{C} \rightarrow \mathcal{C} + \alpha\sum_{m = 1}^{M}|{{W}_m}| + \beta\sum_{m = 1}^{M}|{{W}_m}|^{2}.
\end{equation}
where $\alpha, \beta > 0$.  This cost function is given as a general elastic net regularization ($\alpha \neq 0,~\beta\neq 0$), which accommodates LASSO (least absolute shrinkage and selection operator, $L1:\alpha \neq 0$, $\beta = 0 $) or Ridge ($L2: \beta\neq 0$,~$\alpha = 0$) regularization as limiting cases. We note that the $L1, L2$ regularization terms can be evaluated on a classical computer and added to the cost function evaluated by the quantum computer.

With this general scheme, we can build our hybrid quantum-classical algorithm to find the best parameters $\phi_m,\alpha,\beta$ that minimize the overall cost function. One may tend to implement the popular gradient-based approaches with parameter shifts to search the minima~\cite{Frans}, and this will be very inefficient if the search starts without any prior information about the cost function landscape. 
Instead, we explore a gradient-free algorithm, the Nelder-Mead (NM) optimization algorithm~\cite{Numerical Recipes}, for the cost function to search for global minima.
For current noisy hardware, model training with batched data is plausible. Moreover, with the gradual transition to fault-tolerant hardware, the strategy
with a parallelized NM algorithm with high-performance clusters is still valid with much larger batched data to process before ensemble
averaging. In this case, the gradient-descent-based approach is questionable due to serial
processing and potential barren plateau problems.
We found that convergence to the optimal value of the cost function to high accuracy can typically be found by passing the suboptimal result from the latest global NM search as the warm-start parameters for the next global search iteratively until the desired accuracy is reached.
When model training involves a large amount of training data, training can be broken down into ensemble training with multiple bootstrap data samples in parallel.

\section{Cost function for model training}
By the measurement outcomes from a perfectly trained model with less noisy data and good independent features selected in Eq.~(18),
we expect vanishing measurement results $\mathcal{C} =\langle \hat{M}\rangle = 0$ from a perfect destructive interference, as the predicted response $\hat{y}_{l\in\{0,~1,~2,...,~L-1\}}$ agrees with the actual response $y_{l}$ compared to models that are not well trained.

For a poor model, the model learns the vanishing weights ${\hat W}_{m}$.
If all the response variables are standardized with subtraction from their respective means, we do not consider a constant vanishing bias term, and we expect a finite outcome from Eq.~(18) only contributed from the standardized response data $y^{c}_{l}$ as
\begin{equation}
\mathcal{C}_{0}:=\mathcal{C(\bf W=0)} \propto \sum \limits_{l}(y_l^{c}{\cos\phi_y})^{2}.
\end{equation}

For the worst model, we expect the signs of the weights to be all wrong, and
a much larger probability outcome is expected as
\begin{equation}
\mathcal{C}_{Worst}\propto \sum\limits_{l}(2 y_l^{c}{\cos\phi_y})^{2}.
\end{equation}
We can define the fitness model metrics for the trained model by the coefficient of
determination $R^2$.
We can define $R^2$ as the goodness model metrics for the trained model as $\equiv 1-\mathcal{C}/\mathcal{C}_{0}$ as
\begin{equation}
R^{2}:\left\{
  \begin{tabular}{ll}
  =~$1$~ &~Perfect fitness~ \\
  =~$0$~&~Poor fitness~  \\
  = $-3$~&~Not even a fit~~~~~~~~~~~~~~~~~~~~~~~.
  \end{tabular}
  \right.
\end{equation}
For the worst circumstance where the best estimate of the weight is wrong in sign,
$R^{2}$ could approach the value $-3$. This will be the case where the optimizer
is not set up correctly to find the minima but finds the maxima, or the sign for the rotational angle for the response variable is not well taken off.
For meaningful training results, the goodness metrics should be in the range $R^{2} \in (0,1]$.

Notice that the model metrics $R^{2}$ are independent of any normalization convention in the algorithm.
To conclude whether the measurement result can be differentiated, we can estimate what is needed for the cost
$\mathcal{C}_{0}$ to be resolvable in experiments.
In terms of the standardized response variable $y_{l}^{c}$ after global normalization,
the cost $\mathcal{C}_{0}$
can be expressed as
\begin{equation}
  \mathcal{C}_{0} = \frac{1}{\mathcal{N}}\sum\limits_{l}(y_l^{c}{\cos\phi_y})^{2},
\end{equation}
where $\mathcal{N}$ is the global normalization factor after state preparation.
In terms of standardized classical data by the total variance of training data,
the global normalization factor $\mathcal{N}$ is given by
\begin{equation}
  \mathcal{N} = \sum\limits_{l}{y_{l}^{c}}^{~2} + \sum\limits_{lm}{x_{lm}^{c}}^{2}
  = \sigma_{y^{c}}^2 + \sum\limits_{m}{\sigma^{2}_{x_{m}^{c}}},
\end{equation}
in which the number of rows is given by $L$, the sample variance for the input variable $x_{m}^{c}$ with the column $m$ is given by $\sigma^{2}_{x_{m}^{c}}$, and the sample variance for the response variable $y$ is given by $\sigma^{2}_{y^{S}}$.
Finally, we can arrive at the following expression for the cost $\mathcal{C}_{0}$ as
\begin{equation}
\begin{aligned}
 & \mathcal{C}_{0} = \cos^{2}{\phi_y} \frac{\sigma_{y^{c}}^2}{\sigma_{y^{c}}^2+\sum\limits_{m}\sigma_{x_{m}^{c}}^2} \\
 & \le \frac{1}{1 + F},
\end{aligned}
\end{equation}
in which we have the freedom to fix $\phi_{y}:=\phi_{0}$ at $-\pi$ to optimize the gap between a perfect model and a poor model; the total sample variance of all $M$ features is given by $\sum\limits_{m}\sigma_{x_{m}^{c}}^2$, and the relative sample variance ratio factor $F$ between all features and the response is defined by $F \equiv \sum\limits_{m}
{\sigma_{x^{c}_{m}}^{2}}/{\sigma^{2}_{y^{c}}}=M\bar{\sigma}^{2}_{x^{c}}/\sigma^{2}_{y^{c}}=M\mathcal{V} \in (0,\infty)$, in which $\bar{\sigma}^{2}_{x}:=\frac{1}{M}\sum_{m}\sigma_{x_{m}^{c}}^{2}$ is the average variance per feature excluding the response variable, and $\mathcal{V}$ is the ratio of average variance per feature over the response variable.

For the relative variance factor $F$, it scales with the number of encoded features $M$ under the same standardization procedure before quantum data encoding.
Therefore, we expect the cost function for poor regression training $\mathcal{C}_{0}=\cos^{2}{\phi_{0}}\frac{1}{1+F}$ to scale inversely to the number of features $M$.
For a perfect model to be built, there should be destructive interference that leads to the true cost function ${\mathcal C} = 0$ when the number of shots is large enough (Central Limit Theorem) without physical errors. Ultimately, the variance of the cost function determines the number of shots needed to recover converged results at a high sampling rate. The exact result for one-shot sample complexity per classical optimization iteration is derived in Appendix B. 

\section{End-to-end concrete algorithmic implementation}
\begin{figure}[h]
    \centering
    \includegraphics[width=1.0\textwidth]{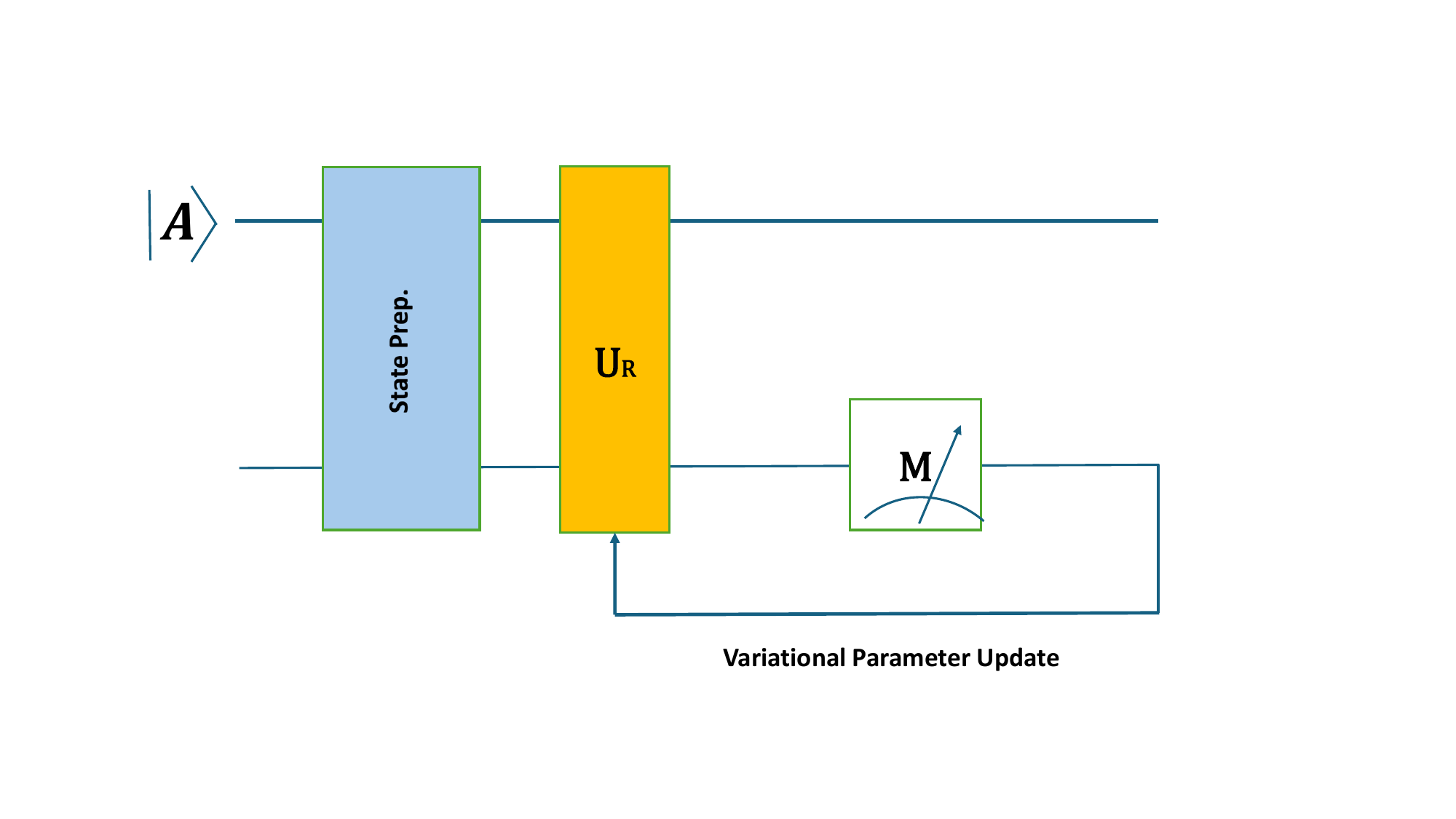}
    \caption{Quantum algorithm modeled by the (quantum) circuit layout}
    \label{fig:A}
\end{figure}

Previously, we described how to implement linear regression in a variational quantum circuit with controlled phase gates. To have an end-to-end solution, we need to consider how to encode the data in the quantum state, that is, how to prepare the data state $|\psi_D \rangle$. To that end, we envision that the state can be prepared using programmable phase gates similar to those used to perform the regression. However, while in the regression step, the phases depended only on the features and were the same for each observation; to encode the data, each distinct phase $\phi_{lm}$ will generally be needed for each data element $x_{lm}$.
We notice that due to the global normalization condition, each normalized element $x_{lm}$ is generally much smaller than 1. This indicates that we can encode these classical training data elements through small phase angles in which $\sin x_{lm} \approx x_{lm}$, that is, the phase angles are approximately the data elements themselves. 
To minimize the potential hardware errors, we prefer low qubit counts while maintaining the simplicity of the algorithm. This is a particularly appealing solution for well-connected and programmable qubits such as Rydberg atom-based and ultra-cold ion quantum platforms~\cite{Small Programmable Cold Ion, Martin, Martin2}.
At this point, we intend to consider the specifics of the data encoding.
For organizational clarity, before diving into detailed gate model implementation, we show the layout of the algorithm first in Fig.~\ref{fig:A} for two different
encoding schemes, in which the functional blocks represent the state preparation, the generation of the quantum regression map, and the decoding through the cost function measurement. 

\subsection{One-hot encoding}

\subsubsection{Quantum data state preparation}

In one-hot encoding, each pair $(l,m)$ is mapped to a single index in $\{0,\ldots, L(M+1)-1\}$ and encoded by the value $1$ in the qubit with the corresponding index.  This would require $L(M+1)$ qubits. One-hot encoding should be avoided for large data sets, as it requires more physical qubits.
For quantum machine learning on near-term quantum devices, this encoding is still useful
for the proof-of-concept of the algorithm we propose, since the circuits to implement it are relatively simple.

With one-hot encoding, the pair $(l,m)$ is mapped to a single index $j = m + l(M+1) \in \{0,\ldots,L(M+1)-1\}$.  The basis state $|lm\rangle$ is then encoded as $|1_j\rangle \equiv |0...010...0\rangle$, which has 1 for qubit $j$ and zero for every other qubit. This leads to
\begin{equation}
    |\psi_{D}\rangle = \sum_{l,m} x_{lm} |lm\rangle
    = \sum_{j=0}^{L(M+1)-1} x_j |1_j\rangle.
\end{equation}
The uniform superposition of a one-hot-encoded state
is the well-known $W$ state that can be prepared by an efficient procedure~\cite{W state}.
The data state $|\psi_D\rangle$ can be prepared using essentially the same procedure but with modified rotation angles to produce the nonuniform amplitudes $x_j$.
The basic building block of this procedure is the gadget
\begin{figure}
\begin{center}
\begin{tikzcd}
& \ctrl{1} & \gate{X} & \qw \\
& \gate{R_{y}(\theta)} & \ctrl{-1} & \qw \\
\end{tikzcd}
\end{center}
\caption{State preparation gadget for one-hot encoding \label{fig:one_hot_gadget}}
\end{figure}
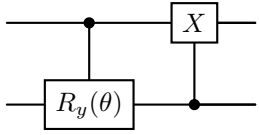
in Fig. 1, consisting of a controlled $\rm{Y}$ rotation followed by a controlled-NOT (CNOT) gate.
Such a gate can be implemented with ease on many experimental platforms (see, for example, Fig. 2 of the reference ~\cite{Small Programmable Cold Ion}).
This gadget maps $|10\rangle$ to $\cos\theta |10\rangle + \sin\theta|01\rangle$.
Starting with the state $|1_0\rangle = |10...0\rangle$ and applying this gadget with various angles to qubit pairs $(0,1), (1,2), (2,3), \ldots$ one can prepare an arbitrary superposition of basis states $|1_0\rangle, \ldots, |1_{L(M+1)-1}\rangle$. For the digital two-local gates, the gate complexity scales as $LM$ in the encoding.

\subsubsection{Quantum regression map}
In the one-hot encoding, the ancilla-controlled phase gate used to impart regression coefficients takes the form
\begin{align}
    U_{C}^{j}(\phi_j) &= \exp\left( -i \phi_j |1\rangle \langle 1| \otimes |1_j\rangle \langle 1_j| \right) \\
    & = \exp\left( -i \phi_j \frac{\rm{Z_{A}-I_{A}}}{2} \otimes \frac{\rm{Z_{j}-I_{j}}}{2} \right)
\end{align}
where $A$ denotes the ancillary qubit, $j$ indexes a data register qubit, $\rm{Z_A}$ ($\rm{Z_j}$) is the Pauli $\rm{Z}$ operator on qubit $A$ ($j$). It can be verified that $U_{C}^{j}(\phi_j)$ yields the desired effect on $x_j$ as
$U_{C}^{j}(|1\rangle \otimes |1_j\rangle) = \exp(-i\phi_j)(|1\rangle \otimes |1_j\rangle)$ and for every other state $U_{C}$ acts as the identity. In Table~\ref{table:full algorithm one-hot}, we summarize the full algorithm before measurement. The time complexity for the regression map is $\mathcal{O}(LM)$ for local gates, but can be further improved with non-local gates $\mathcal{O}(M)$ similar to state preparation (See Appendix C).

\subsubsection{Measurement}
The measurement operator $\hat{M}$ is the summation of individual operators of the form $|lm\rangle \langle lm'|$.
This may be understood as a transition from $j=(l,m)$ to $j'=(l,m')$ which can be achieved by operators of the form $ \rm{S_{j'}^{+} S_{j}^{-}}$ where $\rm{S_j^{+}} = |1\rangle \langle 0| = (\rm{X_j - i Y_j})/2$ is the raising operator on qubit $j$ and $\rm{S_j^{-}} = |0\rangle \langle 1| = (\rm{X_j + i Y_j})/2$ is the lowering operator on qubit $j$. $\hat{M}$ can be written in terms of measurable quantities, as
\begin{align}
\hat{M} &= \sum_{l}\sum_{m, m'}|lm\rangle\langle lm'| \\
 &= \rm{I} + \sum_{l}\sum_{m \ne m'} |lm\rangle\langle lm'| \\
& = \rm{I} + \sum_{l} \sum_{(j<k) = l(M+1)}^{l(M+1)+M}  \left( \rm{S_{j}^{+}S_{k}^{-} + S_{k}^{+}S_{j}^{-}} \right) \\
& = \rm{I} + \sum_{l} \sum_{(j<k)= l(M+1)}^{l(M+1)+M} \left( \rm{X_{j}X_{k}+Y_{j}Y_{k}} \right),
\end{align}
where $\rm{I}$ denotes the global identity (idle) operator. Thus, $\hat{M}$ can be measured as a linear combination of $\rm{I}$, $\rm{X_j X_k}$, and $\rm{Y_j Y_k}$ measurements. 
Taking a two-by-two data table as an example, the basis states for $l=0$, $m=0,1$ are $|1000\rangle$ and $|0100\rangle$.
For $l=1$, $m=0,1$, the states are $|0010\rangle$ and $|0001\rangle$.
For $l=0$ the relevant state transition operators $\rm{{S_{0}}^{+}{S_{1}}^{-}}$ and $\rm{{S_{1}}^{+}{S_{0}}^{-}}$ are given by $|1000\rangle\langle 0100|$ and $|0100\rangle\langle 1000|$. For $l= 1$ the relevant state transition operators $\rm{{S_{2}}^{+}{S_{3}}^{-}}$ and $\rm{{S_{2}}^{+}{S_{3}}^{-}}$ are given by
$|0010\rangle\langle 0001|$ and $|0001\rangle\langle 0010|$.

Since the number of local operators $X_{j}X_{k}, Y _ {j} Y _ {k} $ scales poorly with the number of qubits, we need a more effective measurement scheme
for the cost function $\mathcal{C}=\langle{\hat M}\rangle$.
Shadow tomography has been developed to save the cost of performing the measurements effectively~\cite{Robert}(see Appendix E). 
The one-hot amplitude encoding introduced so far can be resource-intensive in qubits and error-prone, since the number of qubits scales with the number of classical data entries.
However, the overall quantum algorithm is relatively simple but can be noise-prone for noisy quantum hardware due to the number of physical qubits involved. For near-fault-tolerant hardware, we expect that one-hot encoding can be realized.  To mitigate hardware noise and reduce the size of quantum circuits needed, a batch training strategy is employed: First, we divide the training data into numerous batches of smaller bootstrap samples and train a regression model for each batch separately. Then the ensemble-averaged coefficients of the separate regression models are used to produce the ensemble model, as will be demonstrated numerically in Section~{\ref{sec: numerics}}.

\begin{table}
\scalebox{0.85}{
\fbox{
\begin{tabular}{l}
{\bf Multilinear regression model with One-hot Encoding} \\
\toprule
{\bf Input Data}
Globally standardized data matrix $x_{j}\in \mathbb{R}$ \\
{\bf Initialization}\\
$i=0$ (Initialization of the time index $i=0$);\\
$\theta_{j}^{i=0}$ (Initialization of variational parameters);\\
\bf {While ($i < i_{max}$):} \\
\ \ \ {\bf Begin}\\
\ \ \ \ \ \ $\bullet$ The data registry state $|\psi_{D}(i)\rangle$ is initialized as $W$ state with one-hot encoding (Sec. 4.1.1 ); \\
\ \ \ \ \ \ $\bullet$ Initialization of the ancillary state $|+\rangle$; \\
\ \ \ \ \ \ $\bullet$ Application of multi-controlled phase gates with $\prod U_{C}^{j}(i)$ (Sec. 4.1.2);\\
\ \ \ \ \ \ $\bullet$ Application of a Hadamard gate $H$ to the ancillary qubit $A$ (Step 4, Sec. 2.4);\\
\ \ \ \ \ \ $\bullet$ Projective measurement conditioned on the ancillary state $|0\rangle$ (Step 6, Sec. 2.4);\\
\ \ \ {\bf End}\\
\ \ \ {\bf Intermediate Result}: cost function estimation $\langle\hat{M}^{i}\rangle$ (Sec. 4.1.3);\\
\ \ \ {\bf Update}: $\theta_{j}^{i} \rightarrow \theta_{j}^{i+1}$ (classical optimization) \\
{\bf Result} \\
\ \ \  Estimated optimal parameters $\hat{\theta}_{j}=\theta_{j}^{i_{max}}$, optimal weights $\hat{w}_{j}=-\cos(\theta_{j}^{i_{max}})/\cos(\theta_{0}^{i_{max}})$, and the optimal cost function $\langle\hat{M}^{i_{max}}\rangle$.              
\end{tabular}
}
}
\caption{Summary of Quantum Regression Algorithm with One-hot Encoding}
\label{table:full algorithm one-hot}
\end{table}

\subsection{Compact binary encoding}

For the one-hot amplitude encoding, the number of physical qubits allocated to support the information grows linearly with the number of data entries.
To extend the quantum algorithm to current noisy hardware, we need a much more compact encoding scheme to minimize hardware noise due to a much larger qubit count for the same task.
Meanwhile, we want to again keep the structure of the classical data table and use the simple ancilla-controlled phase gates.

For this encoding scheme, the information from the rows and columns is stored in separate qubit registers $|lm\rangle =|l\rangle \otimes |m\rangle$. $l$ and $m$ are encoded in binary using $N_L = \lceil \log_2 L \rceil$ qubits for $l$ and $N_M = \lceil \log_2 (M+1) \rceil$ qubits for $m$. That is, $\ket{l} = \ket{l_{N_L}} \otimes \cdots \ket{l_1}$ and $\ket{m} = \ket{m_{N_M}} \cdots \ket{m_1}$. Take, for example, a $4\times 4$ data table representing 4 observations each of 3 features and 1 response variable. In this case, the row indices $l=0,1,2,3$ would be represented by the four basis states $|00\rangle$,~$|01\rangle$,~$|10\rangle$, and $|11\rangle$, respectively; the same four basis states in the column register would encode $m=0,1,2,3$.

Thus, the number of qubits needed to store the entire data table is approximately $\log_2 L + \log_2 M = \log_2 LM$, which represents a substantial compression.
However, because of compressed encoding, the procedure to impart the regression coefficients into the quantum state has a much higher complexity in terms of one- and two-qubit operations.
Therefore, we will consider an alternative approach that exploits global entangling analog gates native to the latest cold-ion and Rydberg cold-atom systems.

\begin{table}
\scalebox{0.72}{
\fbox{
\begin{tabular}{l}
{\bf Multilinear Regression Model with Compact Binary Encoding} \\
\toprule
{\bf Input Data}\\
Globally standardized data entry $\theta_{k} \approx \tilde{x}_{k}=(2^{-1}(-1)^{x_{k,1}}+2^{-2}(-1)^{x_{k,2}+\cdots 2^{-N_{P}}(-1)^{x_{k,N_{P}}}})$,\\ 
$ k\in \{0,1,\cdots, K-1=L(M+1)\}, x_{k,1}, x_{k,2},\cdots x_{k,N_{P}}\in \{0,1\}$; \\  

{\bf Initialization}\\
$i=0$ (Initialization of the time index $i=0$);\\
$\theta_{m}^{i=0}$ (Initialization of variational parameters);\\
\bf {While ($i < i_{max}$):} \\
\ \ \ {\bf Begin}\\
\ \ \ \ \ \ \ $\bullet$ State preparation for the ancillary qubit, data register, and QPU register: $\ket{\Psi} = \ket{+} \otimes \left( \frac{1}{\sqrt{K}} \sum_{k} \ket{k}_{QPU} \right)\otimes\ket{\mathbf{X}}_{MEM}$ (Sec. 4.2.1):\\
\ \ \ \ \ \ \ \ \ 1. Data state preparation by multi-controlled phase gates $\prod U_{D}^{k}$ operating on the data register, \\
\ \ \ \ \ \ \ \ \ \ \ \ followed by projective measurement on the ancillary state $|-\rangle$; \\
\ \ \ \ \ \ \ \  \ 2. Unitary rotation of the ancillary qubit from $|-\rangle$ state to $|+\rangle$ state;\\
\ \ \ \ \ \ \ \ \ (additional step to glue state preparation and quantum regression map generation)\\
\ \ \ \ \ \ \ $\bullet$ Quantum regression map generation by multi-controlled phase gates $\prod U_{C}^{m}$ operating on QPU register (Sec. 4.2.2); \\
\ \ \ \ \ \ \ $\bullet$ Application of a Hadamard gate $H$ to the ancillary qubit $A$ (Step 4, Sec. 2.4);\\
\ \ \ \ \ \ \ $\bullet$ Projective measurement on the ancillary state $|0\rangle$ (Step 6, Sec. 2.4);\\
\ \ \ {\bf End}\\
\ \ \ {\bf Intermediate Result}: cost function estimation $\langle \hat{M}^{i}\rangle$ (Sec. 4.2.3);\\
\ \ \ {\bf Update}: $\theta_{j}^{i} \rightarrow \theta_{j}^{i+1}$ (classical optimization) \\
{\bf Result} \\
\ \ \  Estimated optimal parameters $\hat{\theta}_{m}=\theta_{m}^{i_{max}}$, optimal weights $\hat{w}_{m}=-\cos(\theta_{m}^{i_{max}})/\cos(\theta_{0}^{i_{max}})$, and the optimal cost function $\langle \hat{M}^{i_{max}}\rangle$.         
\end{tabular}
}
}
\caption{Summary of Quantum Regression Algorithm with Compact Binary Encoding}
\label{table:full algorithm binary}
\end{table}

\begin{figure}[h]
    \centering
    \includegraphics[scale=0.35]{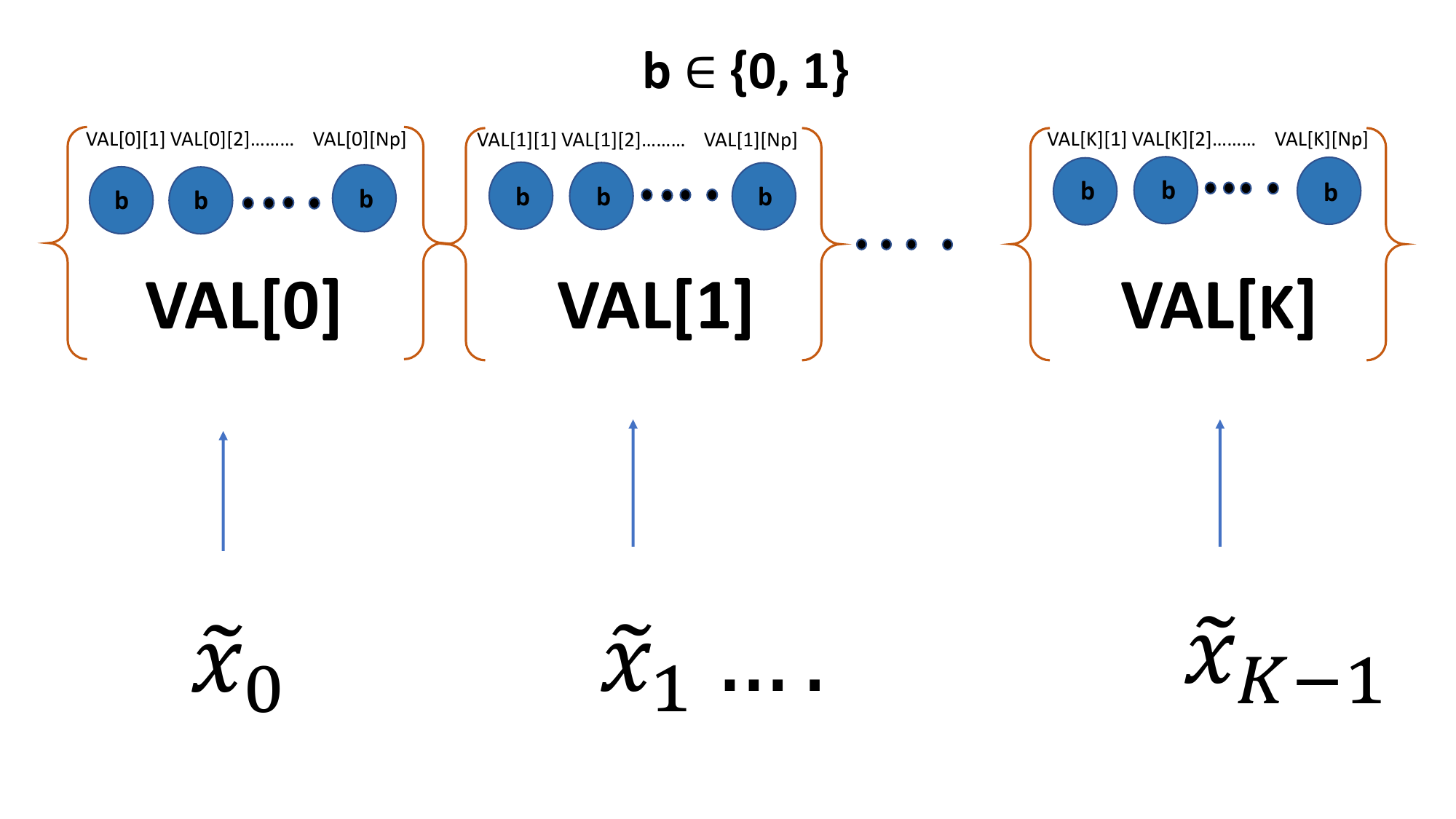}
    \caption{Quantum memory registry resource allocation: a physical qubit is represented by a filled circle
    with a binary code $b \in~\{0,~1\}$.}
    \label{fig:QRA}
\end{figure}

\subsubsection{Quantum data state preparation}

To prepare a quantum state $|\psi_D\rangle$ containing the classical data $\bf X$ in binary strings, we consider the scheme in \cite{Quantum state preparation protocol for encoding classical data into the amplitudes of a quantum information processing register's wave function} in which the real data is first digitized and programmed into a computational basis state supported by the quantum memory register. This needs to be done only once, upfront. Subsequently, each time a copy of the quantum data state $|\psi_D\rangle$ is initialized, it is prepared by a known and efficient circuit that coherently applies the phases stored in the quantum memory register to a quantum processing unit (QPU) register, without destroying the data in the memory register. The quantum information in the memory register can also be reinitialized when the quantum coherence time is surpassed without changing the quantum algorithm in the quantum processing unit.

We assign the index $k=0,\ldots, K-1$ as the entries of a data table, where $K = L(M+1)$.  Each standardized data element $x_k$ is digitized as $\tilde{x}_k$ and stored in a data register (See Fig.~\ref{fig:QRA}):
\begin{equation}
 \ket{\tilde{x}_k}_{VAL[k]}.
\end{equation}
To digitize $x_{k}$, then $x_k \in [-1, 1]$ is approximated using $N_P$ bits of precision as
\begin{equation}
 x_k \approx \tilde{x}_k =  \left( 2^{-1} (-1)^{x_{k,1}} + 2^{-2} (-1)^{x_{k,2}} + \cdots + 2^{-N_P} (-1)^{x_{k,N_P}} \right)
\end{equation}
where $x_{k,1},\ldots,x_{k,N_P} \in \{0,1\}$.  $\tilde{x}_k$ is then stored in the memory as
\begin{equation}
 \ket{\tilde{x}_k}_{VAL[k]} = \ket{x_{k,1}} \ket{x_{k,2}} \cdots \ket{x_{k,N_P}}.
\end{equation}
The full state of the memory register is
\begin{equation}
    \ket{\mathbf{X}}_{MEM} = \bigotimes_{k=0}^{K-1}  \ket{\tilde{x}_k}_{VAL[k]}.
\end{equation}
The total number of qubits for the memory register is $K N_P \approx LM \log_2 (\epsilon^{-1})$ where $\epsilon = 2^{-N_p}$ is the precision of each data element.
While the number of qubits is linear in the size of the data table, as mentioned previously, these qubits need only be kept in a classical digital state.

Once the data has been stored in the memory register, a fixed circuit uses the memory register coherently to ingest the discrete data into the amplitudes of the superposition state $|\psi_D\rangle$ on the QPU register. We introduce an ancillary qubit in the state $\ket{+}$ and an $ N_K$-qubit QPU register in a uniform superposition of all the binary encoded keys, yielding the state
\begin{equation}
     \ket{\Psi} = \ket{+} \otimes \left( \frac{1}{\sqrt{K}} \sum_{k} \ket{k}_{QPU} \right) \otimes\ket{\mathbf{X}}_{MEM},
\end{equation}
in which the state $|k\rangle_{QPU}$ is the shorthand for the encoded key $|l\rangle |m\rangle$ for the data location in the table. Note that the QPU register size $N_K =\log_{2}K \approx \lceil \log_2 (L) \rceil + \lceil \log_2 (M+1) \rceil\approx \log_2 (LM)$ is much smaller than that of the (classical) memory register, which reduces the opportunities for hardware errors.
The essential step is a unitary that transfers the digitized classical data $\ket{\tilde{x}_k}$ to the phase of the ancillary qubit when the key in the QPU register is $k$~\cite{Quantum state preparation protocol for encoding classical data into the amplitudes of a quantum information processing register's wave function}:
\begin{equation}{\label{eq:state_preparation}}
U^{k}_{D} = \exp \left( -i \rm{Z_A} \otimes   \ket{k}\bra{k}_{QPU} \otimes \hat{\theta}_{k} \right)
\end{equation}
The operator $\hat{\theta}_{k}$ is given by
\begin{equation}\label{eq:theta_k}
    \hat{\theta}_k = \sum_{j=1}^{N_P} \Delta \theta_j \rm{Z_{k,j}}
\end{equation}
in which $\rm{Z^{k}_{j}}$ is the operator on the $j$-th qubit on the value register $|VAL\rangle[k]$. When applied to the memory register, $\hat{\theta}_k$ is evaluated to the $N_P$-bit approximation to $x_k$:
\begin{equation}
\hat{\theta}_{k} \ket{\mathbf{X}}_{MEM} = \tilde{x}_k \ket{\mathbf{X}}_{MEM}.
\end{equation}
Notice that the phase $\Delta \theta_j = 2^{-j}$ is predetermined and can be realized by programming quantum gates with suitable gate times and interaction strengths, although fully programmable quantum hardware on a large scale is still an active research area in hardware implementation~\cite{Small Programmable Cold Ion, Monroe}.
The factor $\ket{k}\bra{k}_{QPU}$ in the exponent causes this phase to be produced only if the state of the QPU register matches the key $K$. Explicitly,
\begin{equation}
    U_D^k \left( \ket{b}_A \otimes \ket{k'}_{QPU} \otimes \ket{\mathbf{X}}_{MEM} \right) = e^{-i (-1)^{b} \tilde{x}_k \delta_{k,k'}} \left( \ket{b}_A \otimes \ket{k'}_{QPU} \otimes \ket{\mathbf{X}}_{MEM} \right)
\end{equation}
where $b \in \{0,1\}$. Then
\begin{align}
    \prod_k U_{D}^{k} \ket{\Psi} &=  \left(  \frac{1}{\sqrt{K}} \sum_{k} \frac{ e^{-i \tilde{x}_k } \ket{0}_A + e^{i \tilde{x}_k}  \ket{1}_A }{\sqrt{2}} \otimes \ket{k}_{QPU} \right) \otimes\ket{\mathbf{X}}_{MEM}.
\end{align}
The encoded data state is then realized by projecting the ancillary qubit onto $\ket{-}$:
\begin{align}
    \bra{-}_A \prod_{k} U_D^{k} \ket{\Psi}
    & = \frac{1}{\sqrt K} \sum_{k} \sin \tilde{x}_k \ket{k}_{QPU} \otimes  \ket{\mathbf{X}}_{MEM} \\
    &\approx \ket{\psi_D}_{QPU} \otimes  \ket{\mathbf{X}}_{MEM},
\end{align}
in which the approximation is due to the $N_{p}$-bit approximation of the classical information $x_{k}$ ($\sin \tilde{x}_k \approx \tilde{x}_k \approx x_k$). The success probability observed by the ancillary projection $|\bra{-}_A \prod_{k} U_D^{k} \ket{\Psi}|^2$ is well approximated by
$\frac{1}{K}\sum_{kk'}\tilde{x}_{k}\tilde{x'}_{k'}\delta_{k,k'}=\frac{1}{K}\sum_{k}{\tilde x}_{k}^2$.
(When the $\tilde{x}_{k}$ for each feature is shifted by its mean and scaled optimally as discussed at the end of Appendix B, the success probability scales as $L^{-1}$ as a result of the normalization $\sum_{k}{\tilde x}_{k}^2=(1+M)$, which is fine for noisy quantum hardware and batch training.)

The unitary $U_D = \prod_k U_D^{k}$ is rather complex with terms involving many-qubit Pauli operators. Here we show that we can take advantage of nonlocal M{\o}lmer-S{\o}rensen (MS) gates, which are available in current cold-ion technology~\cite{Peter Zoller, Peter Zoller2} and are an active research area in Rydberg-atom platforms~\cite{Martin, Martin2}. We first expand the key selection operator in terms of Pauli strings and the binary encoding of $k$  as $k = k_{N_K} \cdots k_2 k_1$:
\begin{align}\label{eq:kk expansion}
   \ket{k}\bra{k}_{QPU} &= \prod_{i=1}^{N_K}
   \frac{ \mathbf{1} + (-1)^{k_i}\rm{Z^{QPU}}_{i}}{2} \\
   & = 2^{-N_K} \sum_{P \in \{ \rm{I,Z} \}^{\otimes N_K}} (-1)^{\frak{p}(\rm{P})} \rm{P_{QPU}},
 \end{align}
Here $\frak{p}(\rm{P}) \in \{0,1\}$ is the parity of these bits of $k$ that correspond to factors of $\rm{Z}$ in $\rm{P}$.
As a result,  $U_{D}^{k}$ can be written as $ U_{D}^{k} = \prod_{j=1}^{N_P} U_{D}^{k,j}$ where
\begin{equation} \label{eq:U_kj}
     U_{D}^{k,j} = \prod_{\rm{P} \in \{ \rm{I,Z} \}^{\otimes N_K}}
    e^{-i 2^{-N_{K}} (-1)^{\frak{p}(\rm{P})} \Delta \theta_j \rm{Z_A} \otimes  \rm{P_{QPU}} \otimes \rm{Z_{k,j}}}.
\end{equation}
Notice that each factor in $U_{D}^{k,j}$ is a multi-qubit Pauli rotation, where the operator in the exponent is the tensor product of Pauli $\rm{Z}$ operators operating on selected qubits.
It will soon be possible to implement such rotations efficiently in fully programmable cold ion or cold atom qubit architectures. As discussed in \cite{Peter Zoller2} and Appendix C, a many-qubit rotation can be realized by a short (length $\mathcal{O}(1)$) sequence of nonlocal M{\o}lmer-S{\o}rensen (MS) gates in conjunction with one-qubit ancillary gates on selected qubits. A basic MS gate operation generates a global set of pairwise interactions, while ancillary qubits in conjunction with MS gates generate interactions for Pauli strings on as many qubits as needed. We point out that this is an example of rarely-discussed digital-analog quantum computation~\cite{Eugene}.
In any case, the gate complexity to implement $U_{D}^{k}$ using such an approach is $\mathcal{O}(N_P 2^{N_K}) \approx N_P LM$.
There are $LM$ keys, so the overall gate complexity for state preparation $\prod_{k}U_{D}^{k}$ is $LM N_P 2^{N_K}\approx N_{P}(LM)^2$.
Notice that the state preparation is more demanding due to the quantum data being injected by the quantum memory registry in comparison with the one-hot-encoder introduced earlier, where the classical resources are used.
If we replace the quantum memory registry with the classical resource (classical variable $\theta_{k}$ instead of the quantum variable ${\hat \theta}_{k}$ in Eq.~(33)), the gate complexity of $U_{D}^{k}$ can be further reduced to $\mathcal{O}(LM)$ and the gate complexity for the state preparation is of $\mathcal{O}(L^2M^{2})$. With compilation optimization, it can be reduced to $\mathcal{O}(LM)$~\cite{Frans}.

In comparison, the cost of implementing $U_{D}^{k,j}$ with local digital gates is higher. By the discussion of Hamiltonian simulation on page 210 of \cite{book}, the gate complexity to implement a multi-qubit rotation using local digital gates is roughly proportional to the number of qubits involved. Thus the gate complexity to implement $U_{D}^{k,j}$ using local digital gates is on the order of $\sum_{n=1}^{N_K} n \binom{N_k}{n} = N_K 2^{N_K}$ with
the overall gate complexity for $\prod_{k,j}U_{D}^{k,j}$ estimated as $LM N_{P} N_K 2^{N_K}$. This is greater than the gate complexity of the suggested global MS implementation by a factor of $N_K \approx \log_2 (LM)$.

\subsubsection{{\bf Quantum regression map}}

To impart the regression coefficients into the data state $\ket{\psi_D}$ the memory register is not needed; the coefficients are imparted by the unitary $U_{C}^{m}$, Eq.~(\ref{eq: regression unitary}), acting on the QPU and the ancillary register:
\begin{equation}
    U_{C}^{m} = e^{i\phi_{m} \rm{Z_A} \otimes \mathbf{1} \otimes \ket{m}\bra{m}}.
    \label{eq:reg}
\end{equation}
$U_{C}^{m}$ is analogous to $U_{D}^{m}$ but with two main differences. First, while $U_{D}^{k}$ selects a specific data element $k=(l,m)$, $U_{C}^{m}$ selects only the column $m$ and performs identically on each row $l$ of the data table.  The second difference is that, while the phase imparted by $U_{D}^{k}$ is encoded digitally in the quantum VAL register, the phase appearing in $U_{C}^{m}$ is a simple scalar determined by the regression coefficient.

$U_{C}^{m}$ can be implemented using the same strategy as $U_{D}^{k}$.  Recall that the column index $m$ is represented in binary strings as $m = m_{N_M} \cdots m_2 m_1$ where $N_M = \lceil \log_2 (M+1) \rceil$. Then
\begin{equation}
    \ket{m}\bra{m} = \otimes_{j=1}^{N_M} \ket{m_j}\bra{m_j} = \prod_{j=1}^{N_M} \frac{\mathbf{1} + (-1)^{m_j} \rm{Z_j}}{2}
\end{equation}
where $Z_{j}$ is the operator on $j$-th qubit in the $\ket{m}$ register within the QPU register. Upon factoring, the product $U_C^{m}$ may be written as
\begin{equation}
     U_{C}^{m} = \prod_{\rm{P} \in \{\rm{I,Z}\}^{\otimes N_M}} e^{+i(2^{-N_{M}}\phi_{m} (-1)^{\frak{p}(\rm{P})} \rm{Z_{A}}\otimes {\bf 1} \otimes \rm{P_{QPU}}},
\end{equation}
where this time $\frak{p}(\rm{P})$ is the parity of those bits of $m$  that correspond to factors of $\rm{Z}$ in $\rm{P}$.  Again, these multi-qubit rotations can be implemented either as a multi-qubit controlled gate or using multi-qubit M{\o}lmer-S{\o}rensen gates as discussed above.
The gate complexity for the feature mapping in quantum regression
scales as $2^{N_{M}}\times (M+1) \in  \mathcal{O}(M)$.
In Table~\ref{table:full algorithm binary}, we summarize the full algorithm before the measurement.
In Appendix C, we discuss potential hardware implementation and resources for digital-analog gate operation for interested readers.

\subsubsection{Measurement}
In the compact binary encoding, the measurement operator $\hat{M}$, Eq.~(\ref{eq: measurement operator}) takes a particularly simple form:
\begin{equation}
    \hat{M} = \sum_{l=0}^{L-1} \ket{l} \bra{l}
    \sum_{m,m'=1}^{M} \ket{m} \bra{m'}.
\end{equation}
Using the binary expansion of $\ket{l}$ we have $\sum_l \ket{l} \bra{l} = I^{\otimes N_L}$. Similarly, $\sum_{m,m'} \ket{m} \bra{m'} = (2 \ket{+}\bra{+})^{\otimes N_M} = (I+X)^{\otimes N_M}$. Thus
\begin{align}
    \hat{M} &= 2^{N_M} \rm{I}^{\otimes N_L} \otimes (\ket{+} \bra{+})^{\otimes N_M} \\
    & = \rm{I}^{\otimes N_L} \otimes (\rm{I+X})^{\otimes N_M}
\end{align}
Take a 2-by-4 data table, for example. The $\ket{k}$ states for $l=0$ are $\ket{0}\ket{00}$, $\ket{0}\ket{01}$, $\ket{0}\ket{10}$, and $\ket{0}\ket{11}$; the states for $l=1$ are analogous.  In this case
\begin{align}
    \hat{M} &= \rm{I} \otimes (\rm{I+X}) \otimes (\rm{I+X})
\\
 &= \rm{I \otimes I \otimes I + I \otimes I \otimes X + I \otimes X \otimes I + I \otimes X \otimes X}.
\end{align}
$\ket{\Psi_0} = \psi_{000} \ket{0}\ket{00} + \cdots + \psi_{111} \ket{1} \ket{11}$ denotes the state of the $QPU$ register just prior to measurement.  This reproduces the cost function from all rows as 
\begin{align}
   \mathcal{C} =  \langle \hat{M} \rangle &=
    |\psi_{000} + \psi_{001} + \psi_{010} + \psi_{011} |^2 +
    |\psi_{100} + \psi_{101} + \psi_{110} + \psi_{111} |^2.
\end{align}
In the compact encoding, refer to Appendix E for the practical choice of the shadow tomography measurement protocol and Appendix B for the information-theoretical sample complexity analysis.

\section{Numerical results}
\label{sec: numerics}
Conventionally, to draw reliable interpretations from a trained regression model, we need to characterize the statistics of the uncertainty for the corresponding regression parameter for the predictor variables to justify its relevance in explaining the data. Motivated by the bootstrap aggregation (bagging) and the success of the random forest algorithm, we can build a regression model with bootstrap samples and compute the average and the standard errors (SEs) of the predicted regression coefficients from the ensemble of corresponding regression models by drawing the same number of bootstrap data samples from the original master (data) population (See reference~\cite{bootstrap1, bootstrap2, Statistical Learning} for bootstrap sampling concepts and numerical analysis). The approach is theoretically supported~\cite{BoLasso}. 
Since the qubits in quantum hardware would be noisy when handling a large data set, a plausible solution is to train the regression model from smaller bootstrap samples from smaller subsets of the training data with the same circuit to quantify errors and gather the final bootstrap statistics of the regression parameters by averaging the measurement results from the quantum algorithm running on the quantum hardware. 
Because of the exact mapping of classical regression models into quantum ones in our proposal, the statistical properties for the classical regression model still apply to the hybrid quantum regression model, as
illustrated in the following numerical demonstration as an example. The generalization and extension to other quantum machine models need to be explored further.

Here, we show the promise of quantum-encoded data that can be processed in well-connected quantum hardware and provide an alternative hybrid quantum solution for quantum machine learning applications. For the proposed variational quantum regression (VQR), we show a different and robust strategy to use a global optimization search algorithm to find the optimal regression coefficients to avoid measurement overheads based on gradient-based approaches.
The best estimation can be found by using suboptimal solutions with lower accuracy for regression coefficients as a new ansatz initialization for the next round of the global Nelder-Mead (NM) optimization algorithm until the final converged solution to high accuracy is found.  
For numerical demonstration, we adopt the NM optimization algorithm from SciPy~(an open-source Python library for scientific and technical computing) to validate the batch learning strategy with the analytically known cost function in Eq.~(18) (for larger data applications with distributed bootstrap samples, the distributed NM optimizers can be used).
In the following numerical results, the tuning variables for the cost function are cosine functions for the phase angles
instead of the phase angles.
The search for our optimal solutions is more effective with the new variables because of the unconstrained search
for the NM optimizer. To implement the analogous solution in practice for noisy quantum hardware, one can use NM optimized parameters from the sample mean of bootstrap ensembles via classical simulation as the warm-start condition before applying ensemble-averaged parameter-shifted gradient descent to cut down the informational sample complexity. This procedure is also promising with good classical noise model simulation. 

\subsection{Ensemble model training}
Quantum machine learning from an ensemble model can be useful for statistical modeling, so that it is scalable with large data.
We trained an ensemble model from $N_b$ sets of bootstrap samples of various sizes. The best model is determined by the estimated weight vector ${\bf \widetilde {W}}$ calculated from the estimated feature weights ${\bf {\widetilde W}}=({\widetilde W}_{1}, \widetilde{ W}_{2}, ...., \widetilde{W}_{M})$ from bootstrap samples, that is, ${\widetilde{W}_{i}} = N_{b}^{-1}\sum_{b = 0}^{N_b-1}{\widetilde{W}}_{i}^{b}$ in which $\widetilde{W}_{i}^{b}$ is the weight learned from the batch $b$ for the feature $i$ and the standard errors (SEs) for the weights from model training is denoted as $\delta{\widetilde W}_{i}$.
To validate the ensemble learning, we generate synthetic and standardized classical data sets with a deterministic linear map with small randomness between the $M$ features $X_{j} = (X_{j,1}, X_{j,2},..., X_{j, M})$ and the target variable $Y_{j} \in \mathcal{R}$.
Specifically, the ideal (noiseless) linear map is given by the expression
$Y_{j} =  X_{ji} \hat{W}_{i}$ where $X_{ji}$ is the $L$-by-$M$ data matrix and
the best weight vector $\hat{\bf W} = (\hat{W}_{1},\hat{W}_{2},...,\hat{W}_{M})$ of size $M$ after data standardization.
We let each feature follow the uniform random distribution between values $[-1,1]$ to cover the feature space. 
For the response column $Y_j$, it is generated by the linear map $Y_{j} = {X_{ji}}W_{i}$ with the random variables $\{W_{i = 1,2,.., M}\}$ with the ideal population ${\bf \hat W}=({\hat W}_{1}=1,{\hat W}_{2}=2,{\hat W}_{3}=3,..,{\hat W}_{M}=M)$ and the standard deviation $\delta$(the same for each feature) from its mean value ${\hat W}_{i}$.
We would expect the model training to be more uncertain for the first few features due to the smaller signal-to-noise ratio ${\widetilde W}_{i}/\delta \widetilde{W}_{i}$ where $i \in \{ 1,2,..., M \}$ from an equal number of bootstrap samples with different sample sizes.
Notice that the weights ${\widetilde{W}}_{i}$ and the SEs $\delta\widetilde{{\it W}_{i}}$ from training are in overline tildes to differentiate from the true weight ${\hat W}_{i}$ and the standard deviation $\delta$ from the data generation, respectively.

\begin{table}[h]
\centering
\hspace{0.8cm}
\scalebox{1.127}
{\renewcommand{\arraystretch}{1.0}{
\begin{tabular}{||c|c|c|c|c|c|c|c|c|c||}
\hline
Sample size & \scriptsize ${\widetilde W}_{1}$ & \scriptsize ${\widetilde W}_{2}$ & \scriptsize ${\widetilde W}_{3}$ & \scriptsize ${\widetilde W}_{4}$ & \scriptsize ${\widetilde W}_{5}$ & \scriptsize ${\widetilde W}_{6}$  \\
\hline
10 & 0.99938 & 2.00113 & 2.99967 & 3.99968 & 5.00004 & 6.00009 \\
20 & 1.00008 & 2.00004 & 3.00002 & 4.00001 & 5.00002 & 5.99998 \\
40 & 1.00004 & 2.00007 & 3.00003 & 4.00013 & 5.00000 & 6.00012 \\
60 & 0.99999 & 2.00001 & 2.99998 & 4.00006 & 5.00001 & 6.00004 \\
100 & 1.00001 & 2.00001 & 3.00004 & 4.00002 & 5.00000 & 6.00002 \\
150 & 0.99997 & 2.00001 & 3.00002 & 4.00001 & 5.00004 & 6.00002 \\
\hline
\end{tabular}
}
}
\begin{minipage}{1.0\textwidth}
            \centering
            \includegraphics[width=\linewidth]{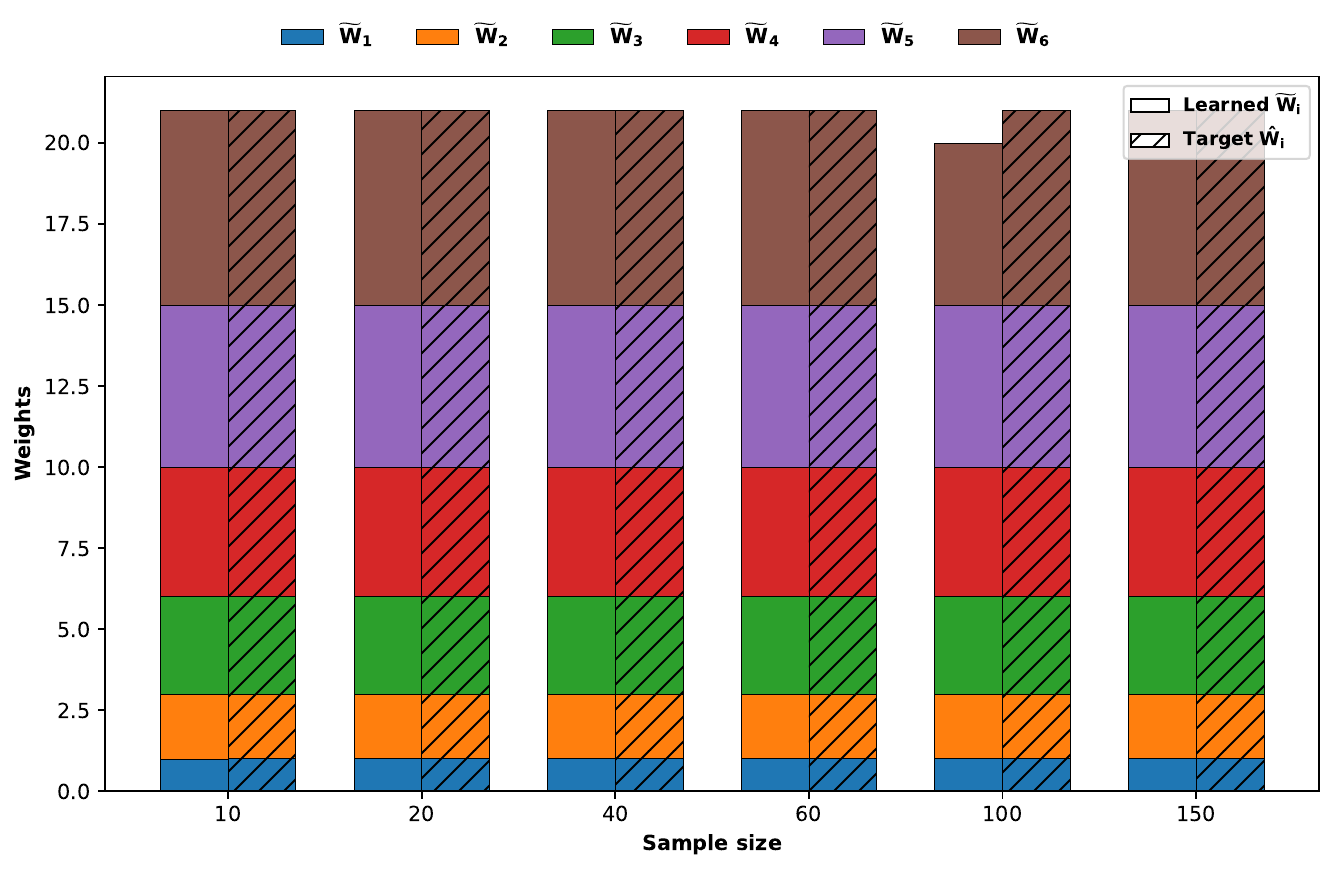}
        \end{minipage} 
\caption{Estimated (learned) weight vectors by sample mean versus bootstrap sample sizes for the noiseless population.
Numerical values of the weight vectors are shown in the top panel. The same pictorial display is shown on the bottom panel with solid-colored texture, where  
features are coded in distinct-colored bar charts as a function of sample size. 
The height of the colored bar represents the magnitude of the feature weights.
The attached (duplicated) bar charts with tilted line texture show the ideal theoretical result: the target weight vector ${\bf \hat W}=({\hat W}_{1}=1,{\hat W}_{2}=2,{\hat W}_{3}=3,..,{\hat W}_{M}=M)$.}
\label{table:weight}
\end{table}

\begin{table}[h]
\begin{center}
\hspace{0.34cm}
\scalebox{1.16}
{\renewcommand{\arraystretch}{1.0}{
\begin{tabular}{||c|c|c|c|c|c|c|c|c||}
\hline
Sample size & \scriptsize $\delta \widetilde{W}_{1}$ & \scriptsize $\delta \widetilde{W}_{2}$ & \scriptsize $\delta \widetilde{W}_{3}$ & \scriptsize $\delta \widetilde{W}_{4}$ & \scriptsize $\delta \widetilde{W}_{5}$ & \scriptsize $\delta \widetilde{W}_{6}$ \\
\hline
10 & 0.02039 & 0.03495 & 0.00725 & 0.01284 & 0.00934 & 0.00781  \\
20 & 0.00204 & 0.00118 & 0.00156 & 0.0016 & 0.00154 & 0.00217
     \\
40 & 0.00153 & 0.00316 & 0.003 & 0.00299 & 0.00155 & 0.00245
    \\
60 &  0.00077 & 0.00072 & 0.00076 & 0.00172 & 0.00056 & 0.00088
    \\
100 & 0.00087 & 0.00104 & 0.00106 & 0.00074 & 0.00092 & 0.00174  \\
150 & 0.00069 & 0.00052 & 0.00049 & 0.00034 & 0.0009 & 0.00079
     \\
\hline
\end{tabular}
}
}
\end{center}
\begin{minipage}{1.0\textwidth}
            \centering
            \includegraphics[width=\linewidth]{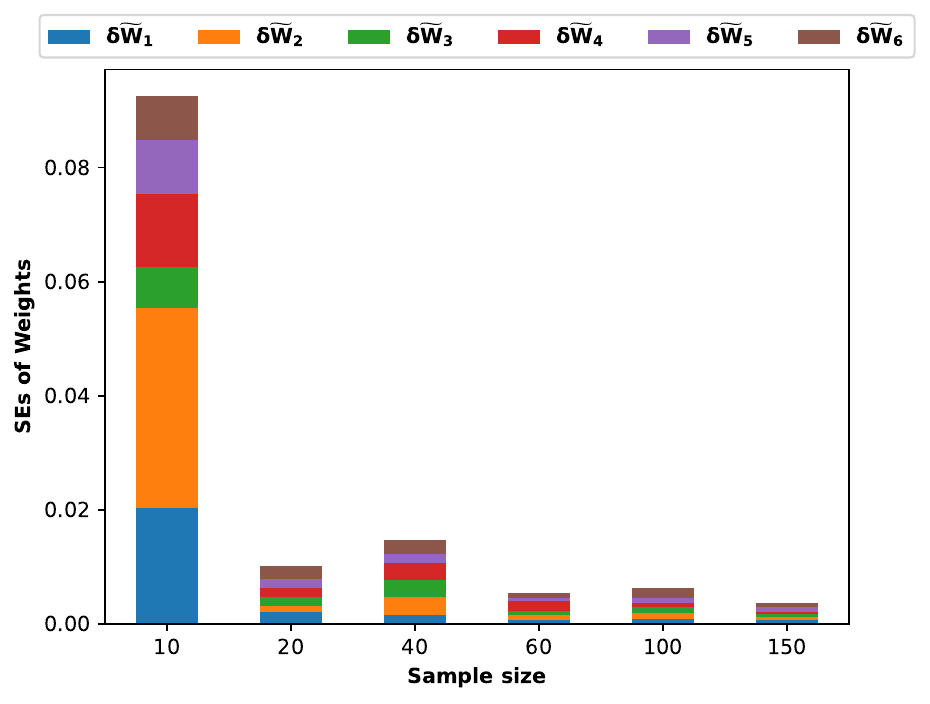}
        \end{minipage} 
\caption{Standard errors (SEs) of features by sample mean versus bootstrap sample sizes for the noiseless population.
Numerical values of the SEs are shown in the top panel. The same pictorial display is shown on the bottom panel with solid-colored texture, where  
features are coded in distinct-colored bar charts as a function of sample size. 
The height of the colored bar represents the SEs of the features.
}
\label{table:var-weight}
\end{table}

\begin{table}[h]
\begin{center}
\scalebox{0.95}
{\renewcommand{\arraystretch}{1.0}{
\begin{tabular}{||c|c|c|c|c|c|c|c|c|c||}
\hline
Sample size & \scriptsize $t_{1}$ & \scriptsize $t_{2}$ & \scriptsize $t_{3}$ & \scriptsize $t_{4}$ & \scriptsize $t_{5}$ & \scriptsize $t_{6}$     \\
\hline
10 & 49.0036 & 57.25731 & 413.59599 & 311.55844 & 535.45297 & 768.20304  \\
20 & 490.03614 & 1697.92281 & 1924.45761 & 2506.22625 & 3237.56853 & 2759.18714 \\
40 & 653.08249 & 632.31813 & 1000.64225 & 1339.80026 & 3227.4301 & 2453.23403  \\
60 & 1291.6086 & 2787.94434 & 3970.08524 & 2321.44326 & 8930.72652 & 6798.62171 \\
100 & 1152.23006 & 1930.15583 & 2824.33706 & 5433.56226 & 5444.65661 & 3455.64034 \\
150 & 1457.42293 & 3855.09323 & 6125.15592 & 11916.92773 & 5580.49053 & 7587.86445 \\
\hline
\end{tabular}
}
}
\end{center}
\begin{minipage}{1.0\textwidth}
            \centering
            \includegraphics[width=\linewidth]{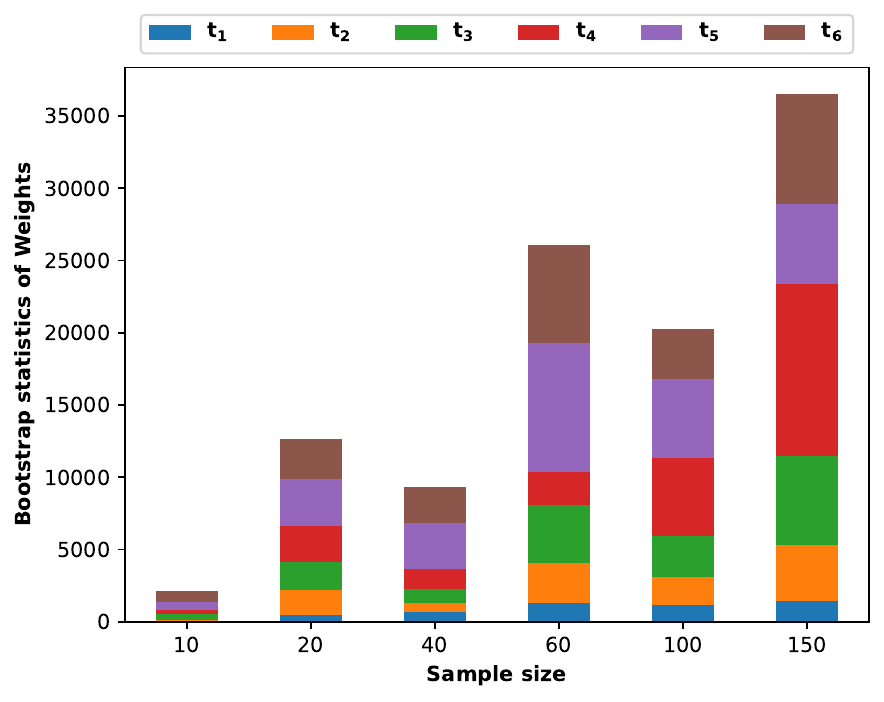}
        \end{minipage} 
\caption{Bootstrap $t$-statistics of features by sample mean versus bootstrap sample sizes for the noiseless population.
Numerical calculation of the $t$ values is shown in the top panel. The same pictorial display is shown on the bottom panel with solid-colored texture, where  
features are coded in distinct-colored bar charts as a function of sample size. 
The height of the colored bar represents the $t$ values of the features.
}
\label{table:t}
\end{table}

A bootstrap sample is a random data set from the original master population with replacement to avoid external correlation induced by the sampling.
For our numerical demo, the master population data set has $N_{b} = 1024$ data records/rows, and we drew 1024 bootstrap samples, each of which has a much smaller chosen sample size. The regression weight vector learned by the proposed regression algorithm from the $1024$ bootstrap re-samples with the respective sample sizes of $10$, $20$, $40$, $60$, $100$, and $150$ records. The zero-bias term is guaranteed to be negligible from data standardization by subtraction from the sample mean.
Including additional columns from the response variable for the quantum encoding, we can emulate the quantum regression training with classical methods
$13$ qubits~($2^{13} = 1024 \times 8$) without padding additional zeros.
For the presentation of numerical results with completeness and clarity in tables, we attach an additional bar chart on the side to assist the readers' comprehension with a visual aid.

Due to the variance of the bootstrap samples, the trained weight vectors fluctuate among these samples.
To establish our baseline errors from sampling and training, we show the ideal case with six features where the training data has no noise $\delta = 0$ to observe if we can emulate the learning.
As shown in Table~\ref{table:weight}, the training reproduces the theoretical values for the synthetic data we generate with the ideal weight vector $\hat{\bf W} = (1,2,3,4,5,6)$, and we observe that the bootstrap sampling for the learning is reproduced for various batch sizes.
SEs of the weight vector $\delta {\bf \widetilde W} = (\delta \widetilde{W}_{1}, ~\delta \widetilde{W}_{2},....,\delta \widetilde{W}_{6})$ stay small and almost unchanged for distinct batch sizes, as shown in Table~\ref{table:var-weight}. With the SEs in weight staying more or less constant, we expect a much larger $t$ for the features with higher weights.
If we look at the $t$-statistics metrics defined by the ratio
$t_{i} = {\widetilde W}_{i}/\delta {\widetilde W}_{i}$ as shown in Table~\ref{table:t}, we observe these values are much greater than one, representing the statistical significance of the learned results.
This shows that bootstrap sampling analysis is a valuable tool and generalizable in practice beyond the Gaussian noise hypothesis~\cite{Statistical Learning}, typically imposed in traditional statistical analysis.

\subsection{Robustness to Gaussian noise}

To further confirm the practicality of the training approach with noise present in the map between features ${X_{i}}$ and the response variable $Y_{i}$, we perform the simulation with the noise level $\delta = 0.1$. For this case, we observe the
deviation of the learned weight vectors away from the ideal case without noise. With small sample sizes $10,20$, and $40$, the sample mean weights can deviate from the theoretical weight vector more than what is indicated by the noise $\delta = 0.1$ in Table~\ref{table:weight-random}.
This is due to the sample variance being more pronounced at smaller batch sizes, as also indicated in the noise-free case $\delta = 0$ in Table~\ref{table:var-weight}.

For larger sample sizes $60$,~$100$, and $150$, we do observe that the sample mean weight vector from training mostly reproduces what is expected for the noise level
$\delta = 0.1$. The SEs of the weight vectors for the noisy cases are shown in Table \ref{table:var-weight-random}. When the learned weight vectors deviate significantly from theoretical values, we observe a corresponding larger SE for the weight vector. This correlation guides how reliable our learned weight vectors are.
For example, for the first feature $\widetilde{W}_{1}$ with the sample size $20$, we see a large deviation from the theoretical value $1\pm 0.1$.
We also observe a larger deviation in its SE: $\delta \widetilde W_{1}$ at the sample size $20$.
In Table \ref{table:t-random}, we observe that the overall $t$ values are lower in comparison with the cases with no noise $\delta = 0$ due to the presence of non-sampling noise. In addition, we can identify that the overall $t$ values are the largest for the batch size $150$. This indicates that we
can use bootstrap sampling with the optimal sample size of $150$.

For larger batch sizes greater than $150$ (not shown), we start to observe deviations from what we expect from theoretical values for the weight vector $\bf \hat W$.
This is because the ensemble training from the resampled data sets is underfitting due to higher chances of duplicated data records in each sample, leading to training bias.
Even though the bias hinders us from drawing quantitative inferences from the data,
this behavior does not prevent us from selecting important features based on the $t$-statistics metrics defined by the ratio
$t_{i} = \widetilde{W}_{i}/\delta {\widetilde W}_{i}$ as shown in Table~\ref{table:t-random} and can be avoided with smaller bootstrap sample size. Note that this is also the case when there is no noise $\delta = 0$~(Table~\ref{table:t}), but it occurs at a larger batch size $> 150$, not shown in Table~\ref{table:weight}.

\begin{table}[h]
\begin{center}
\hspace{0.7cm}
\scalebox{1.124}
{\renewcommand{\arraystretch}{1.0}{
\begin{tabular}{||c|c|c|c|c|c|c|c|c|c||}
\hline
Sample size & \scriptsize ${\widetilde W}_{1}$ & \scriptsize ${\widetilde W}_{2}$ & \scriptsize ${\widetilde W}_{3}$ & \scriptsize ${\widetilde W}_{4}$ & \scriptsize ${\widetilde W}_{5}$ & \scriptsize ${\widetilde W}_{6}$  \\
\hline
10 &  1.24838 & 2.01457 & 2.92003 & 3.85253 & 4.69864 & 5.62674\\
20 & 0.02961 & 2.66497 & 1.64403 & 3.86877 & 5.15871 & 5.44711  \\
40 & 1.80006 & 0.74388 & 2.9371 & 3.97106 & 5.45333 & 5.65369\\
60 & 0.89376 & 1.83072 & 2.97385 & 4.07048 & 4.74884 & 5.72274 \\
100 & 0.63928 & 1.19575 & 3.04848 & 3.61821 & 5.09804 & 6.29805 \\
150 &  1.14361 & 1.79783 & 2.8523 & 3.95466 & 4.75087 & 5.82112 \\
\hline
\end{tabular}
}
}
\end{center}
\begin{minipage}{1.0\textwidth}
            \centering
            \includegraphics[width=\linewidth]{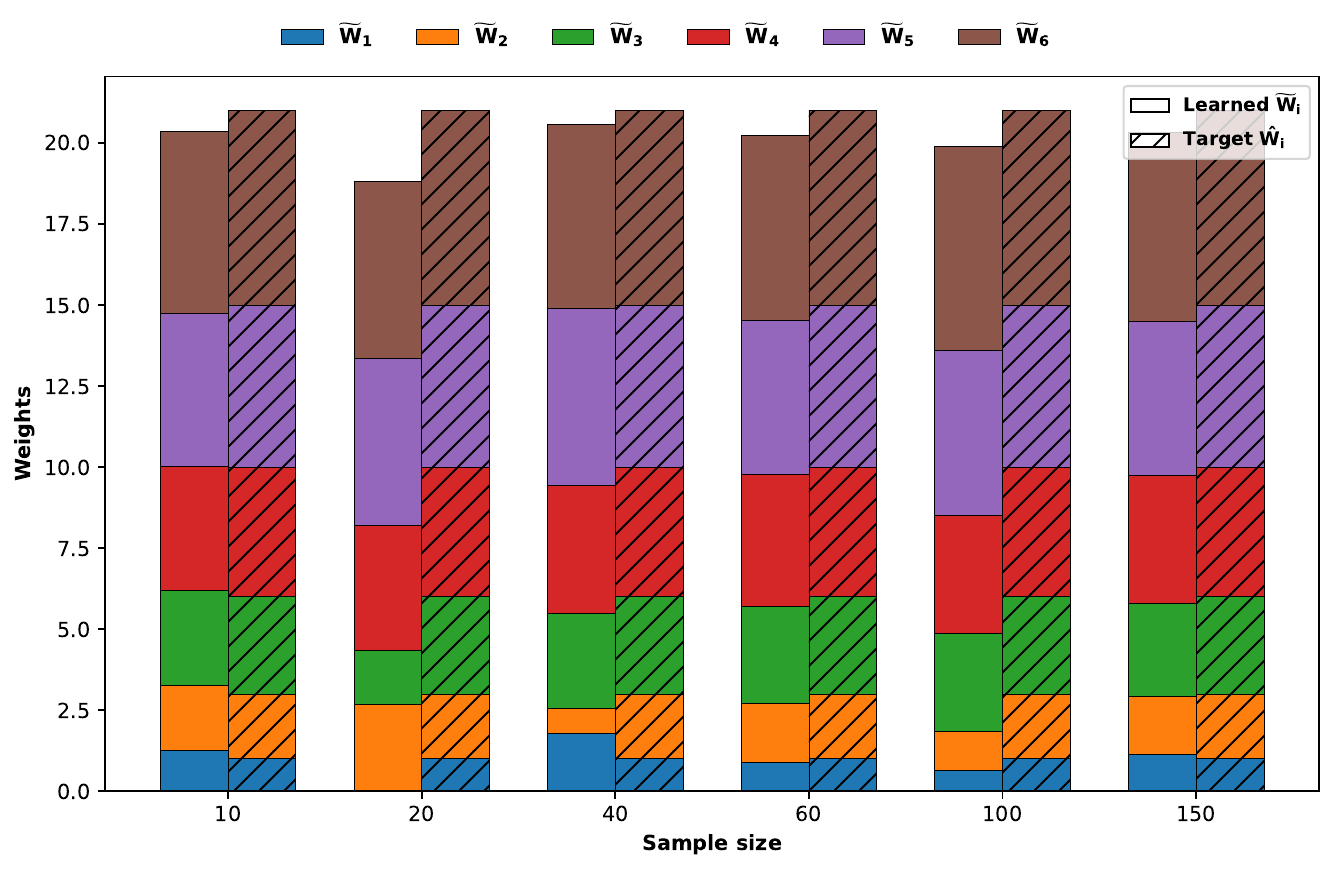}
        \end{minipage} 
\caption{Weight vectors by sample mean versus bootstrap sample size with noisy population $\delta = 0.1$, generated by the noisy regression map ${\bf W}=\hat{\bf W}[1+\mathcal{N}(0,\delta^2)]$, in which $\mathcal{N}(0, \delta^2)$ is the Gaussian (normal) noise distribution with the standard deviation $\delta = 0.1$. The attached (duplicated) bar charts with tilted line texture show the ideal theoretical result: the target weight vector ${\bf \hat W}=({\hat W}_{1}=1,{\hat W}_{2}=2,{\hat W}_{3}=3,..,{\hat W}_{M}=M)$.
}
\label{table:weight-random}
\end{table}

\begin{table}[h]
\begin{center}
\scalebox{1.1}
{\renewcommand{\arraystretch}{1.0}{
\begin{tabular}{||c|c|c|c|c|c|c|c|c||}
\hline
Sample size & \scriptsize $\delta \widetilde{W}_{1}$ & \scriptsize $\delta \widetilde{W}_{2}$ & \scriptsize $\delta \widetilde{W}_{3}$ & \scriptsize $\delta \widetilde{W}_{4}$ & \scriptsize $\delta \widetilde{W}_{5}$ & \scriptsize $\delta \widetilde{W}_{6}$   \\
\hline
10 &  10.3005 & 12.11829 & 13.45395 & 10.91457 & 7.85075 & 15.71212 \\
20 &  18.87171 & 28.18201 & 32.64924 & 14.18558 & 18.51711 & 21.57881 \\
40 & 16.54549 & 46.15104 & 6.43802 & 17.95377 & 20.97624 & 8.10855 \\
60 & 5.92399 & 5.39804 & 7.86823 & 6.12054 & 7.67524 & 5.49687 \\
100 & 9.85255 & 17.82527 & 8.84548 & 19.16585 & 10.84676 & 9.2562 \\
150 & 3.92479 & 3.14805 & 2.88643 & 3.30615 & 3.54041 & 3.47353 \\
\hline
\end{tabular}
}
}
\end{center}
\begin{minipage}{1.0\textwidth}
            \centering
            \includegraphics[width=\linewidth]{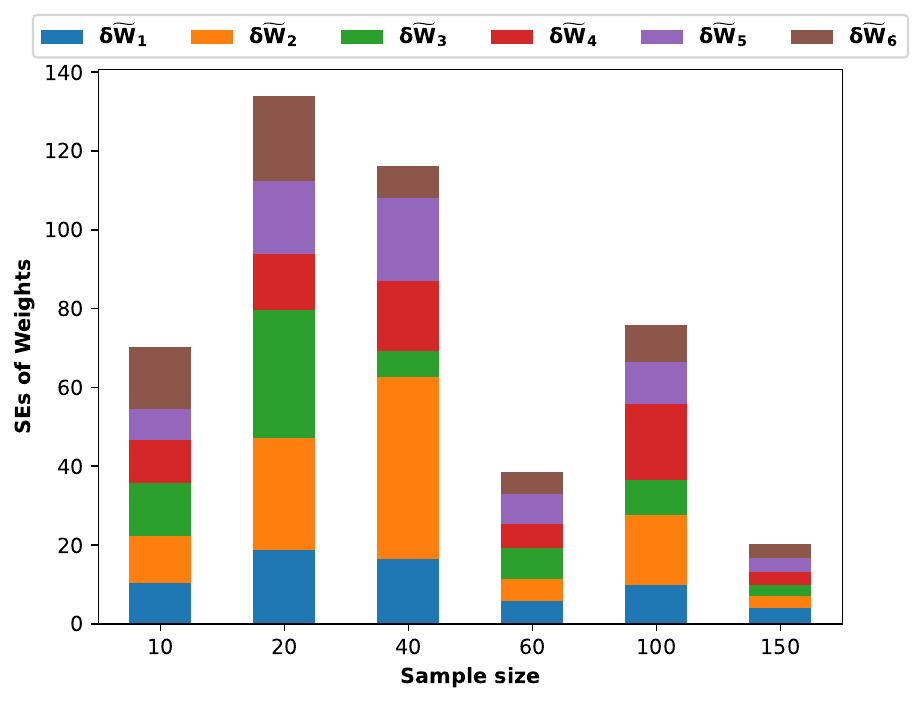}
        \end{minipage} 
\caption{SEs of the weight vectors by sample mean versus bootstrap sample size with the noisy standard deviation $\delta = 0.1$.}
\label{table:var-weight-random}
\end{table}

\begin{table}[h]
\begin{center}
\hspace{0.8cm}
\scalebox{1.115}
{\renewcommand{\arraystretch}{1.0}{
\begin{tabular}{||c|c|c|c|c|c|c|c|c|c||}
\hline
Sample size & \scriptsize $t_{1}$ & \scriptsize $t_{2}$ & \scriptsize $t_{3}$ & \scriptsize $t_{4}$ & \scriptsize $t_{5}$ & \scriptsize $t_{6}$     \\
\hline
10 &  0.1212 & 0.16624 & 0.21704 & 0.35297 & 0.5985 & 0.35811 \\
20 &  0.00157 & 0.09456 & 0.05035 & 0.27273 & 0.27859 & 0.25243 \\
40 & 0.10879 & 0.01612 & 0.45621 & 0.22118 & 0.25998 & 0.69725 \\
60 & 0.15087 & 0.33915 & 0.37796 & 0.66505 & 0.61872 & 1.04109 \\
100 &  0.06488 & 0.06708 & 0.34464 & 0.18878 & 0.47001 & 0.68041 \\
150 &  0.29138 & 0.57109 & 0.98818 & 1.19615 & 1.3419 & 1.67585\\
\hline
\end{tabular}
}
}
\end{center}
\begin{minipage}{1.0\textwidth}
            \centering
            \includegraphics[width=\linewidth]{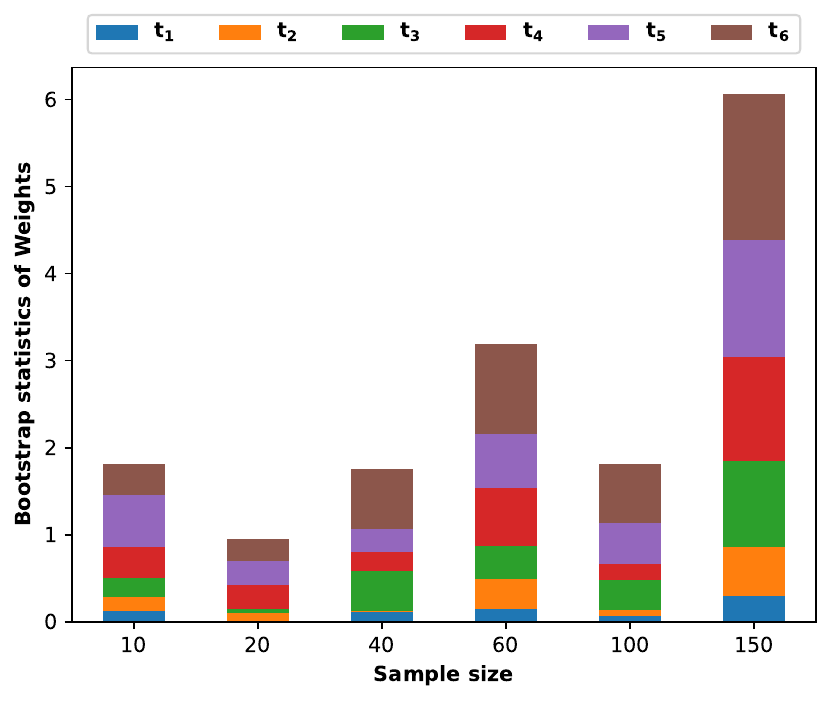}
        \end{minipage} 
\caption{Bootstrap $t$-statistics for the weight vectors by sample mean versus bootstrap sample size with the noisy standard deviation $\delta = 0.1$.}
\label{table:t-random}
\end{table}

\subsection{Feature importance and regularization}
In machine learning, we may have a potentially large list of features that can be used to describe the mapping between the response variable and input variables. Regularization provides an algorithmic way to select the optimal subset of original features quickly before performing a more detailed bootstrap sampling analysis for the finalized features. Regularization
penalizes models with many features with important weights to avoid overfitting the noise present in the training data and to induce more error in noisy hardware.
Since the regularization is done outside the quantum loop, this is a valid strategy for hybrid quantum machine learning. Here, we demonstrate that the optimal feature selection can be enabled by turning on regularization in the cost function.

\subsection{Nonlinear map extension}
\begin{figure}[h]
    \centering
    \includegraphics[scale=0.85]{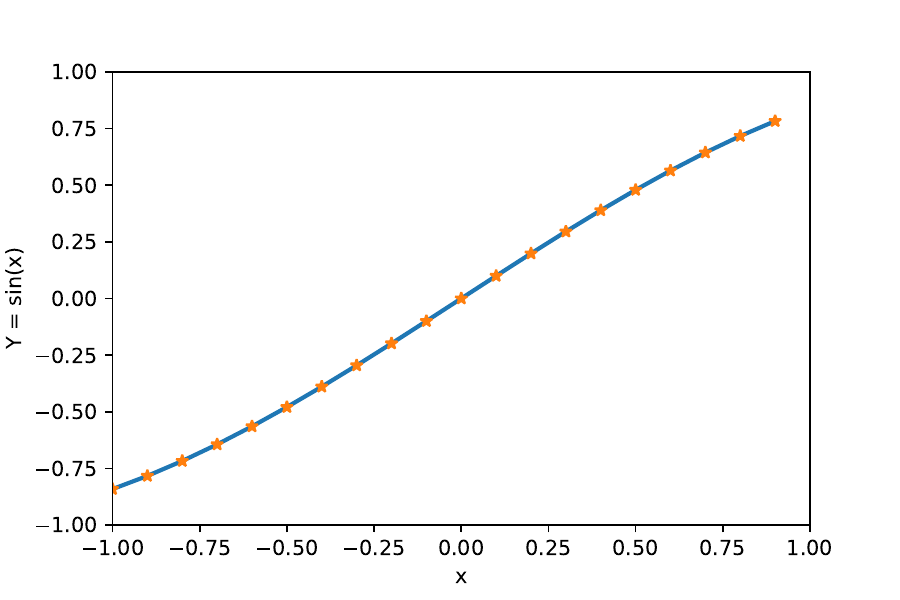}
    \caption{
    Demonstration of nonlinear regression through classical polynomial feature expansion and successful feature selection
    by $L1$ regularization.
    Prediction agreement with the weights from theory results (solid line) and the weights from training from simulation (stars) are truncated to the fourth decimal accuracy: \\
    ${\bf {\widetilde W}}=(0.9992, 0.000001435, -0.1629, -0.000002908, 0.004134, -0.0004442, 0.000004760, \\ 0.0000012079, 0.000005189, 0.0003221, 0.0008932,0.0001738,0.000001149, -0.000001869, \\ -000001375)$.
    }
    \label{fig:agreement}
\end{figure}

To establish the baseline, we generate synthetic data without Gaussian noise $\delta_{0} = 0$ where the response variable $Y=\sin( x)$ depends on the independent real variable $x$ in infinite order and the values are distributed randomly between the values $[-1,1]$. This is an infinite series for any real $x$ values, but can be truncated to a finite series when $x$ ranges between $[-1,1]$, which is the case for our normalized features. For regularization, we test with $L1$ regularization and $L2$ regularization.
We found that the $L1$ regularization works robustly with the NM optimization.
In the following demonstration, we show that the nonlinear features can be built first in the feature space so that the linear regression algorithm can be used for building the nonlinear regression model by feature preprocessing.
We generate the synthetic data with controlled mapping between predictors~(features) ${\bf X} = (X_{1} = x, X_{2} = x^{2}, X_{3} = x^{3} ..., X_{15} = x^{15})$ and the target~(response) variable $Y \in \mathbb{R} = \sin( x) = \sum_{n \in +\mathbb{Z}} (-1)^{n+1}{x}^{2n-1}/(2n-1)!$. The number of population records
$2^5$ are generated where each feature $X_{n}$ is uniformly generated between values $[-1,1]$. With $L1$ regularization, we use the very small
regularization parameter $\alpha = 1.2 \times 10^{-7}$, and alternating signs for the initial weight ansatz as $\sign(W_{1}, W_{3}, W_{5} .., W_{15}) = (+1, -1, +1,.., -1)$.
Our hybrid algorithm converges to the optimal weight parameter ${\bf \widetilde W} \approx {\bf \hat W}$, which selects the first few odd terms as the important features.
Notice that the optimization may end up with a much smaller cost function, but with the wrong signs.
However, this confirms the experience that constraints from domain knowledge are typically required since the classical optimization algorithms can only be used as a filter for possible solutions, and even a mathematical global minimum solution may not be reasonable for the domain of applications.
What we found is that $L1$ regularization works with the proper regularization parameters $\alpha$. The number of vanishing weights in the converged weight vector reveals itself to select the feature effectively.
Taking the best-learned weight to produce the predicted value $Y$ for $x$ values ranging between $[-1, 1]$ as shown in Fig.~\ref{fig:agreement}, we can reproduce what we expected from synthetic training data.

\section{Conclusion and discussion}
We present an inherently explainable quantum regression algorithm and detail the plausible
implementation with quantum hardware with more connectivity. The unitary transformation by the algorithm is constructed without
theoretical errors from any unitary gate decomposition, such as Trotter errors or random circuit ansätze. The model weights responsible for the explainability of the regression model are derived and identified as
the cosine function of the variational rotation angles via controlled phase gates. We expedite the
realization in different hardware through the concrete discussion in hardware implementation (which can be migrated to more modern versions of similar platforms through compilation), typically
ignored in other algorithm work, by showing how the state vector evolution can be determined
by controlled phase gates with high connectivity precisely. The variational regression algorithm is most
likely to demonstrate optimal gate complexity in hardware with highly connected neutral cold atoms or cold ion systems with fewer errors, despite being realized first via a superconducting IQM machine with 20 qubits with a non-synthetic dataset~\cite{Frans}.

Although amplitude-encoded non-uniform state preparation is still costly at large data limits
(see Appendix D), we mitigate this problem by replacing master data training with bootstrap samples (known to be useful beyond Gaussian probability distributions) and a distributed ensemble
regression model training from each sample, as demonstrated in the numerical results and recent experimental results (with gradient descent optimization) by the IQM machine~\cite{Frans}. This does not pose the notorious barren
plateau problem on average sense by the weak gate control noise $\delta\phi_{m}\ll \phi_{m}$ from the gate parameter $\phi_{m}$ over-rotation or under-rotation (since the cost function gradient $|\partial \mathcal{C}/\partial \phi_{m}|$ is proportional to $|\sin{2(\phi_{m} +\delta\phi_{m})}|\approx |\sin{2\phi_m}\langle \cos{2\delta \phi_{m}}\rangle|=|\sin{2\phi_{m}}|\exp{(-2\sigma_{Noise}^{2})}$ where $\sigma_{Noise}^{2}\ll 1$ is the weak noise variance from zero mean). 

In addition, we show that a quantum regression map with the compressed encoded data structure
can greatly reduce the gate complexity compared to the classically constructed regression map.
The compressed encoding strategy potentially mitigates hardware noises at different levels, including incoherent errors
and read-out errors (Appendix B), which are particularly detrimental in solid-state-based platforms.
By further improving the current algorithm's measurement blueprint, the readout errors can be substantially reduced using a cost function derived from the probability of a clean ancillary qubit, and the associated algorithm complexity analysis will be one of the top priorities for the extended work to evaluate potential quantum advantages. Continued support from funding agencies is vital to accelerate these discoveries at scale.

We would like to emphasize the important role of digital-analog gates (with optimal fidelity via control calibration) in reducing gate complexity exponentially when producing a regression map. This exemplifies the constructive feedback loop between desirable quantum hardware capabilities and end-to-end quantum algorithms with utility. Due to the superiority of the compressed encoding and comparable data state preparation complexity per shot,
it is potentially feasible to improve the algorithm toward a larger scale from our $20$ qubit experimental realization in an IQM machine. For larger data state preparation with a near-fault-tolerant machine, we envision
the possibility of avoiding state preparation early in the algorithm by postselection to trim the end-to-end shot complexity
for the algorithm, in addition to the replacement of classical optimization with a quantum optimization primitive.  

Furthermore, we point out the implementation of nonlinear regression models with a quantum linear regression
map but with preprocessed (classical) columns of nonlinear features, as shown in the last instance in the numerical results. This reduces the circuit depth during model training and is useful even beyond noisy quantum hardware. This shares the gist of the Carleman linearization for solving nonlinear differential equations quantumly.

To conclude, we formulate an inherently explainable hybrid quantum regression algorithm for near-term fault-tolerant or noisy quantum hardware with a data-aware information-theoretical bound for the shot sample complexity for a classical optimization iteration identical to that via shadow tomography up to a prefactor. The potential applications and implications from this inherently interpretable and explainable construction with inputs from diverse domains of interest in quantum machine learning will be exciting~\cite{Katlin}. The constrained interpretable framework will make the quantum circuit design more structural and potentially lead to quantum algorithm speedup.    
It would be desirable to implement, benchmark, and experiment with optimal compilations and diverse quantum hardware
with merits to further scrutinize the quantum utilities.
We expose the important nuances for training hybrid variational regression models with inherent explainability and interpretability.
We hope our work will broaden the view of interpretable quantum machine learning for science and industrial applications where the understanding of models
matters most.

\section{Acknowledgments}
We acknowledge Phil Lotshaw for his candid feedback on the original manuscript.  C.-C. Joseph Wang acknowledges Ryan Bennink for his cordial discussion and financial support for the project through the DOE Office of Science, Office of ASCR, under FWP No. ERKJ354.

\appendix
\section*{Appendix}
\section{Proof of cost function from measurement}
\begin{equation}
\begin{aligned}
& |\Psi_{0}\rangle = \sum_{l',~m'} x_{l'm'}\cos{\phi_{m'}}|l'm'\rangle, \\
& \hat{M} = \sum_{l''}\sum_{m''}\sum_{m'''}|l''m''\rangle\langle l''m'''|=\hat{I}+\sum_{l}\sum_{m\neq m'}|lm\rangle\langle lm'|,\\
& \left(\sum_{m\neq m'}:~shorthand~notation~for \sum_{m}\sum_{m'\neq m}, m \in \mathcal{Z}:\{0,1,2,\cdots,M\}, M \ge 1\right)\\
& \hat{M}^{2}=\hat{I}+2\sum_{l}\sum_{m\neq m'}|lm\rangle\langle lm'|+(\sum_{l}\sum_{m\neq m'}|lm\rangle\langle lm'|)^{2},\\
& =\hat{I}+2(\hat{M}-\hat{I})+M\hat{I}+(1-\delta_{M,1})\times(\sum_{l}\sum_{m\neq m'}|lm\rangle\langle lm'|)\\
& = \hat{I}+2(\hat{M}-\hat{I})+M\hat{I}+(1-\delta_{M,1})(\hat{M}-\hat{I})= (3-\delta_{M,1})\hat{M}+(M-2+\delta_{M,1})\hat{I},\\
& \hat{M}|\Psi_{0}\rangle = \sum_{l'}\sum_{l''}\sum_{m'}\sum_{m''}\sum_{m'''} x_{l'm'}\cos{\phi_{m'}}\langle l''m'''|l'm'\rangle |l''m''\rangle, \\
& = \sum_{l'}\sum_{m'}\sum_{m''}x_{l'm'}\cos{\phi_{m'}}|l'm''\rangle. \\
&~\rm Using~the~orthogonality~relation, \\
& \langle l''m'''|l'm'\rangle = \delta_{l''l'}\delta_{m'''m'} \\
& \langle \Psi_{0}| \hat{M} |\Psi_{0}\rangle =
\sum_{ll'}\sum_{m,~m',~m''} x_{lm}x_{l'm'}\cos{\phi_m}\cos{\phi_{m'}}\langle lm|l'm''\rangle \\
& = \sum_{l}\sum_{m}\sum_{m'} x_{lm}x_{lm'}\cos{\phi_{m}}\cos{\phi_{m'}} \\
& = \sum_{l}(\sum_{m} x_{lm}\cos{\phi_{m}})^{2}, \\
&~\rm Q.E.D.
\end{aligned}
\label{eq:cost function}
\end{equation}

\section{One-shot sampling complexity}
With physical noise errors, the ideal theoretical result will be shifted.
Notice that the error $\delta \epsilon$ scales differently for different encoding schemes. For example,
assuming identical read-out error $\delta_{M} > 0$ for a physical qubit states '$0$','$1$', 
the read-out operator ${\hat R}_{0}$ for the '$0$' state is ${\hat R}_{0}= (1-\delta_{M})|0\rangle\langle 0|+ \delta_{M}|1\rangle\langle 1|$. 
For '$1$' state, the read-out error operator is $\hat R_{1}= \delta_{M}|0\rangle\langle 0| + (1-\delta_{M})|1\rangle\langle 1|$. Therefore, we can estimate the net encoded readout error in leading order in $\delta_{M}$
for one-hot encoding with $N_Q=L(M+1)$ physical qubits as
$\langle \Psi_{0}|{\hat M}\otimes ({\hat R}_{0}^{\otimes N_{Q}-1} \otimes {\hat R}_{1})|\Psi_{0}\rangle=(1-N_{Q}\delta_{M}) \times \langle \Psi_{0}|{\hat M}|\Psi_{0}\rangle=(1-\delta \epsilon(L,~M))\times \mathcal{C} $, in which the minus sign represents the reduction in the cost function. Similarly, for compressed encoding,
the error is exponentially suppressed to leading order in $\delta_{M}$ as given by
the expression $\langle \Psi_{0}|{\hat M}\otimes({\sum_{q,q'}\hat R}_{0}^{\otimes q} \otimes {\hat R}_{1}^{\otimes N_{L}-q}\otimes {\hat R}_{0}^{\otimes q'} \otimes {\hat R}_{1}^{\otimes N_{M}-q'})|\Psi_{0}\rangle = [1-(N_{L}+N_{Q})\delta_{M}] \times \langle \Psi_{0}|{\hat M}|\Psi_{0}\rangle = [1 -\delta \epsilon(L,~M)] \times \mathcal{C}$, in which $N_L = \lceil \log_2 L \rceil$ physical qubits and $N_M = \lceil \log_2 (M+1) \rceil$ physical qubits are used. We recapitulate the results in the abstract.

The sample complexity can be estimated by Bernstein's inequality as follows. When the deviation condition 
$|\mathbb{E}[\hat{M}]-\langle \hat{M} \rangle| \le \epsilon$ with confidence $1-\alpha$ is met 
between the sample mean  $\langle\hat{M}\rangle$ and the true mean $\langle\hat{M}\rangle$, the probability of the deviation can be bounded by ${\rm Pr}[|\mathbb{E}[\hat{M}]-\langle \hat{M}\rangle|\ge \epsilon]\le \alpha=\exp(-N_{shot}\epsilon^2/2\sigma^2)$; therefore, the number of shots $N_{{\rm shot}}$ required is given by $\mathcal{O}\left(\sigma^2/\epsilon^2)\ln(1/\alpha)\right)$ where the true variance $\sigma^{2}=\langle\hat{M}^2\rangle-\langle \hat{M}\rangle^{2}=(3-\delta_{M,1})\mathcal{C}-\mathcal{C}^2+(M-2+\delta_{M,1})$ is defined, and the mean $\langle \hat{M}^2 \rangle= (3-\delta_{M,1})\langle \hat{M}\rangle+(M-2+\delta_{M,1}\hat{I})$ can be derived due to the fact that $\hat{M}^2=(M-2+\delta_{M,1})\hat{I}+(3-\delta_{M,1})\hat{M}$ (See Appendix A). Therefore, when one has access to the cost function measurement $\hat{M}$ in one shot, we arrive at the cost function-dependent sample complexity  $N_{\rm shot} \approx \mathcal{O}\left(\sigma^2(\mathcal{C})\right)
 \frac{\ln{(1/\alpha)}}{\epsilon^2}=\mathcal{O}\left((3\mathcal{C}-\mathcal{C}^2+M-1)\frac{\ln{(1/\alpha)}}{\epsilon^2}\right)$ in which $\mathcal{C} \in [0,\mathcal{C}_{0}]\times [1-\delta\epsilon(L,M)] $ and $\mathcal{C}_{0}=1/(1+M\mathcal{V})$.
 In current practice, we have no access to quantum information with all measurements in one shot. Therefore, shadow tomography was proposed, and we can replace the shot complexity variance $\sigma^{2}$ with the shadow norm depending on the shadow tomography protocols~\cite{Robert}. The corresponding shadow norm is discussed in Appendix E.

For the largest gap separation between $\mathcal{C}_{0}=\cos^{2}{\phi_{0}}\frac{1}{1+M}=\frac{1}{1+M}$ and $\mathcal{C}=0$, the classical data can be rescaled in each column to be identical, leading to $\mathcal{V}\approx 1$ and then we can perform global normalization with the summation of the resulting variance. The global normalization does not change the feature weights, but the rescaled variance for each column does change beforehand.
With the rescaled classical information, we can recover the classical regression coefficient of the original features
$W_{m}^{O}$ before standardization by the identity $\hat{W}_{m}^{c} = \hat{W}_{m}^{O}|\sigma_{x}^{O}|/|\sigma_{y}^{O}|$.

To extract the row-local cost function $\mathcal{C}({\bf W})$ by projected measurement $\langle M \rangle$ with a gradient descent optimizer or parameter shift optimization, one can encounter barren plateaus~\cite{Barren plateau} in the cost function landscape. The gradient $\frac{\partial {\rm C}}{\partial \phi_{m}}=2\sin{\phi_{m}}\sum_{m'}\sum_{l}x_{lm}x_{lm'}\cos{\phi_{m'}}$ for a feature $m$
, in which the quantum amplitude contraction $\sum_{l}x_{lm}x_{lm'}$
 is equal to the bounded Pearson's correlation coefficients $\sum_{l}x_{lm}^{c}x_{lm'}^{c}$ from the classical data with the superscript $c$, $|r_{m,m'}|=|\sum_{l}x_{lm}^{c}x_{lm'}^{c}|\in [0,1]$, divided by the global normalization scaling factor $\mathcal{N}$. 
The global normalization factor $\mathcal{N}$ is proportional to $(M+1)$ after a typical classical data standardization with mean subtraction and identical standard deviation division for all variables before data uploading (see Appendix B).
Barren plateaus exist ($\frac{\partial \mathcal{C}}{\partial \phi_{m}} \rightarrow 0$) only with a large number of features (wide data table) $M \gg 1$, causing overfitting regardless of whether all the independent features considered are perfectly correlated with the response. (When the training data are noisy with negligible correlation with the response variable, noise-induced barren plateaus exist, consistent with the existence of bad training data containing no correlation pattern with the response.)  
The hybrid algorithm does not encounter the typical barren-plateau problem associated with a finite number of features considered during training, thanks to the row-local cost function we constructed. In the presence of noise, we can perform better model training using batched data from resampling data to mitigate error issues from measurement or cross-talk between physical qubits in the encoding, and potentially improve parameter-shift optimization in the presence of the noise-induced barren plateau by noisy training data.

\section{Hardware implementation}
In cold-ion hardware, native gates include arbitrary one-qubit Pauli rotational gates and the two-qubit $\rm{XX}$ gate~\cite{Small Programmable Cold Ion}. The controlled phase (CPH) gate and the CNOT gates can be realized using the $XX$ gate in conjunction with Pauli one-qubit rotations. The $H$ gate can be decomposed as $\rm{{R_{X}}(\pi){ R_{Y}}(\pi/2)}$ also on the platform. (Note that arbitrary one-qubit gates plus any entangling two-qubit gate constitute a universal gate set.)

In the Rydberg atom platform~\cite{Rydberg1}, any rotation in the Bloch sphere can be implemented, and the native two-qubit gate is $\rm{ZZ}$ type. The CNOT gate can also be decomposed in this platform as $\rm{(I \bigotimes H) C_{Z} ( I \bigotimes H)}$
where the controlled $\rm{Z}$ gate $\rm {C_{Z}}$, which is a special case for the controlled phase gate, is enabled by Rydberg states. The $\rm{CPH}$ gate can be decomposed in principle in terms of the $\rm{CNOT}$ gate with a one-qubit rotational gate~\cite{book}.
\subsection{One-hot encoding}
The specifics of the $\rm{CPH}$ gates vary with the encoding schemes.
For the one-hot amplitude encoding, the factorization of controlled two-body Pauli rotations $U_{C}^{j}$ along an axis is required.
Typically, this can be achieved in a preferred Pauli $\rm{Z}$ axis in a particular platform up to
a single-qubit rotation from a native axis to the $\rm{Z}$ axis. For example, the native axis for cold ions would be Pauli $\rm{X}$
and the native axis for the Rydberg atom will be Pauli $\rm{Z}$.
The multi-qubit controlled phase gate can be implemented for the native Pauli $\rm{Z}$ axis as
$U_{C}^{m}=\rm{\prod_{j}\exp(-i\phi_{m}\frac{Z_{A}-I_{A}}{2}\otimes\frac{Z_{j}-I_{j}}{2})}$.
Equivalently, it can also be decomposed locally as
\begin{equation}
    U_{C}^{m}=\rm{\prod_{j}e^{-i\frac{\phi_{m}}{4} Z_{A}\otimes Z_{j}}
    e^{+i\frac{\phi_{m}}{4} Z_{A}\otimes I_{j}}
    e^{+i\frac{\phi_{m}}{4} I_{A}\otimes Z_{j}}
    e^{-i\frac{\phi_{m}}{4} I_{A}\otimes I_{j}}},
\end{equation}
in which the last unitary exponential factor is the idler unitary operator. It is state-independent and can be dropped. By digital 1-local and 2-local gate operations, the gate complexity for each feature is $\mathcal{O}(L)$. With $M$ features, the gate complexity will be $\mathcal{O}(LM)$, the same as the gate complexity with the state preparation as discussed in Section 3.~1.~1.
For partial globally addressed analog gate operation $U_{C}^{m}$ on each feature,
the gate complexity is greatly reduced ($\mathcal{O}(1)$) and scales as $\mathcal{O}(M)$ in total. With a programmable, fully connected global analog gate with all features, the involved commutative Pauli operators,
$\prod_{m} U_{C}^{m}$ can be fused into a global unitary, and therefore, the gate complexity can be minimized to
$\mathcal{O}(1)$ at the expense of the gate complexity $\mathcal{O}(M)$ for the classical controls.
For the hardware with native $\rm{X}$ axis such as cold ions, we need to apply Pauli $\rm{Y}$ rotation $\rm{R_{Y_{A, j}}(-\pi/2)}$ to each physical qubit state in $U_{C}(\phi_{m})$ as
\begin{equation}
\begin{aligned}
&  U_{C}^{m} = \prod_{j} \rm{R_{Y_{A}}(+\pi/2)} \rm{R_{Y_{j}}(+\pi/2)} e^{-i\frac{\phi_{m}}{4} \rm{X_{A}\otimes X_{j}}}\rm{R_{Y_{j}}(-\pi/2)}\rm{R_{Y_{A}}(-\pi/2)}~\\
& \otimes \rm{R_{Y_{A}}(+\pi/2)}  e^{+i\frac{\phi_{m}}{4} \rm{X_{A}\otimes I_{j}}} \rm{R_{Y_{A}}(-\pi/2)} \\
& \otimes e^{+i\frac{\phi_{m}}{4} \rm{I_{A}\otimes X_{j}}},
 \end{aligned}
\end{equation}
in which the digital and local decompositions have arrived.
For the state preparation, the gate needs to be applied to each qubit reserved for the data registry.

\subsection{Compact binary encoding}
For the compact binary encoder, controlled phase gates are more complicated to implement. We suggest the application of a global entanglement gate, the M{\o}lmer-S{\o}rensen ($\rm{MS}$) gate with an ancillary qubit, to achieve the quantum logic gate~\cite{Peter Zoller}.
The $\rm{MS}$ gate unitary operator in trapped cold ions is typically expressed as
\begin{equation}
    U_{MS}(\theta_{MS},\phi)=\rm{\exp(-i\frac{\theta_{MS}}{4}(\cos(\phi)S_{X}+\sin(\phi)S_{Y})^{2})},
\end{equation}
in which $\rm{S_{X,Y}=\sum_{i}{X_{i},\sum_{i} Y_{i}}}$ are the collective Pauli operators.
With the help of an ancillary qubit, Pauli-string operation along a Pauli axis $\rm{X, Y}$ can be enabled by choosing
the value and the sign of the phase $\phi$ to be ${0,\pi}$.
For $\phi =(0,\pi)$, $U_{MS} = (U_{MS}^{X}(\theta_{MS},\phi=0), U_{MS}^{Y}(\theta_{MS},\phi=\pi))$.
Angle $\theta_{MS}$ can be used to tune the rotation angle for the MS gate, where the exhaustive implementation was listed in reference to Table $1$~\cite{Peter Zoller}. Notice that any discrepancy between the algorithms we develop can be easily adjusted to the native preferred axis for any platform after global rotation without many difficulties. The same comments apply to the one-hot encoder for state preparation with controlled-phase gates based on the MS gate. For Rydberg atoms, research on $\rm{MS}$ gates for two-atom and multi-atom systems is still in its infancy~\cite{Martin, Martin2, Lukin}.

For well-connected qubits, the multi-controlled phase gates would be sufficient to implement with only one ancillary qubit in principle, especially when the connectivity is close to infinitely long-ranged. The practicality of the implementation is limited by the range of the phase gates that can be applied uniformly across physical qubits.
For example, for cold ions in a one-dimensional linear Paul trap and a two-dimensional Penning trap,
the ions are mostly uniformly distributed with long-range Ising interactions at the center of the trap. Therefore, it would be wise to select ions away from the trap edges as the data register to reduce the complexity of waveform engineering. Due to this controlled scalability limitation, we anticipate that machine learning will be limited to a certain number of qubits, which in turn limits the amount of the batched training data that can be encoded for concurrent training.
\section{Resource estimation}
Our gate complexity analysis from the previous sections (Appendix C and Section 3) indicates that digital global gates reduce the gate complexity compared to digital local gates.
For the one-hot encoder, with available local and global gates, the gate complexity is $T_{O} \in \mathcal{O}(LM+M)$ ($\mathcal{O}(LM)$ for the state preparation and the reduced complexity $\mathcal{O}(M)$ from the quantum regression map is from the encoded data table structure (see Appendix C) $\mathcal{O}(LM)$ in terms of the number of physical qubits.
For the compact binary encoder, the gate complexity is $T_{C} \in \mathcal{O}(L^2M^2+M)$, including state preparation and regression map generation, respectively.

To decide the shot cost for the encoders, we also need to consider factors beyond gate complexity, such as the product of the qubit footprint (space complexity) and gate complexity. For the one-hot encoder $Q_{O}\in \mathcal{O}(LM)$, it is exponentially more costly than the cost function for the compact binary encoder
$Q_{C} \in \mathcal{O}(\log_{2}LM)$.
Since we need to have a tall table ($L \gg M+1$), in which $M+1$ is the estimate for the Vapnik–Chervonenkis dimension for the linear regression model,  with a generalizable linear regression model without overfitting in statistical learning theory, 
the shot cost ratio $R_{C}/R_{O} = T_{C}Q_{C}/T_{O}Q_{O} \in \mathcal{O}(\log_{2}LM)$.
For classical algorithms, the widely accepted gate complexity of the regression map is $\mathcal{O}(LM^2)$, which is inferior to that of the quantum regression map with the embedded data structure.

However, the gate complexity for classical algorithms for classical data preparation is $\mathcal{O}(LM)$, the same as the one-hot encoder
for the corresponding quantum data state preparation, but better than the compact binary encoder $\mathcal{O}(L^2M^2)$ without compilation optimization for the circuit, since there is no structural map to take advantage of. With a global compilation optimization, the quantum state preparation for the compact encoder can be improved to $\mathcal{O}(LM)$~\cite{Frans}.

The total resource cost $R_{CL}$ for the classical computation that includes the memory cost $\mathcal{O}(LM)$ by physical bits for
the classical data would be $R_{CL} \in \mathcal{O}(L^{2}M^{3})$, which shows a slight disadvantage over the physical qubit resource cost $R_{O}\in \mathcal{O}(L^2M^2)$
and $R_{C} \in \mathcal{O}(LM\log_{2}{LM})$ after compilation optimization. The classical optimization is the common resource cost for both the hybrid quantum algorithm and its classical algorithm to compare.

\section{Cost function shadow tomography measurement protocol}
\subsection{One-hot encoding}
For one-hot encoding, the measurement operator $\hat{M}$ requires estimating $\mathcal{O}(LM)$ terms of the form $X_jX_k$ and $Y_jY_k$ across all row-local pairs. Rather than measuring each term independently, which scales linearly with the number
of observables, classical shadows~\cite{Robert} offer a dramatic reduction in
measurement cost. By applying random global Clifford circuits to the post-selected state $|\Psi_0\rangle$ and recording the computational-basis outcomes, a classical shadow of size $\mathcal{O}(\log M)$ suffices to simultaneously predict all $M$ expectation values with high probability. Random global Clifford measurements are the natural choice here for two reasons. First, the observables $\hat{M}_{jk} = X_jX_k + Y_jY_k$ are non-local, spanning pairs of qubits across the full $n = L(M+1)$qubit register, which means their locality $k$ is not small and the shadow norm, which is proportional to the observable variance under random Pauli measurements, would scale as $4^k$, becoming prohibitively large. 
Random Clifford measurements instead bound the shadow norm through the Hilbert-Schmidt norm, which remains $\mathcal{O}(1)$ for the normalized observables in $\hat{M}$, yielding a system-size-independent sample complexity. For the empirical estimation of the shadow norm using the $3$-design
property of the Clifford group (refer to Proposition~1 of~\cite{Robert}).

\subsection{Compact binary encoding}
In the compact binary encoding, $\hat{M} = I^{\otimes N_L} \otimes (I + X)^{\otimes N_M}$ decomposes into a sum of $2^{N_M}$ tensor-product Pauli observables acting on the $N_M$ column-register qubits, where $N_M = \lceil\log_2(M+1)\rceil$. While this is logarithmically smaller than in the one-hot case, the need to estimate multiple Pauli strings simultaneously remains. Classical shadows based on random single-qubit Pauli measurements are the natural choice in this setting, for two complementary
reasons. First, the compact binary encoding is specifically designed to minimize the qubit footprint, making hardware simplicity a priority; random Pauli measurements require only independent single-qubit Clifford rotations and no entangling gates, in contrast to the global Clifford circuits needed in the one-hot case. Second, the observables in $\hat{M}$ act nontrivially only on the $k = N_M$ column-register qubits and are therefore genuinely local. Under random Pauli measurements, the shadow norm scales as $4^k$, which depends only on this small locality $k = N_M = \lceil\log_2(M+1)\rceil$ and not on the total qubit count $N_L + N_M$.
The locality of the tensor-product structure allows the shadow norm to be compressed to the $k$-qubit support of each observable, and a Pauli-basis expansion followed by a case-by-case evaluation of the single-qubit Clifford averages via the $3$-design property yields (Proposition~3 of~\cite{Robert})
to be $\mathcal{O}\left(\log(2^{N_M}) \cdot 4^{N_M}/\varepsilon^2\right)$ therefore suffices to predict all terms in $\hat{M}$ simultaneously to accuracy $\varepsilon$.

\end{document}